\documentclass[]{IEEEtran}
\usepackage{mathrsfs}
\usepackage{amsfonts}
\usepackage{graphicx,cite,epsfig,amssymb,amsmath}
\usepackage{color,xcolor}
\usepackage{pifont}
\usepackage{stmaryrd}
\usepackage{setspace}
\usepackage{subfigure}
\usepackage{cite}
\usepackage{array}
\usepackage{float}
\usepackage{verbatim}
\usepackage{multirow}
\usepackage[linesnumbered, ruled]{algorithm2e}
\usepackage{epstopdf}
\usepackage{mathtools}
\usepackage{diagbox}
\usepackage{cases}
\usepackage{makecell}
\usepackage{booktabs}

\providecommand{\algorithmname}{Algorithm}
\floatname{algorithm}{\protect\algorithmname}
\newcommand{\bm}[1]{\mbox{\boldmath{$#1$}}}
\newcommand\emptyDiag[2][]{\diagbox[innerwidth=\widthof{#2}, height=\line, #1]{}{}}
\newtheorem{thm}{Theorem}

\newtheorem{lem}{Lemma}

\usepackage{arydshln}
\allowdisplaybreaks[4]
\begin{document}

\title{RIS-Assisted Communication Radar Coexistence: Joint Beamforming Design and Analysis}
\author{\IEEEauthorblockN{Yinghui He, \IEEEmembership{Student Member,~IEEE,} Yunlong Cai, \IEEEmembership{Senior Member,~IEEE,} \\Hao Mao, and Guanding Yu, \IEEEmembership{Senior Member,~IEEE}}\\
\thanks{Manuscript received August 23, 2021; revised December 10, 2021; accepted January 14, 2022. The work of G. Yu was supported in part by GDNRC[2021]32. The work of Y. Cai was supported in part by the National Natural Science Foundation of China under Grants 61971376 and 61831004, the Zhejiang Provincial Natural Science Foundation for Distinguished Young Scholars under Grant LR19F010002. (\emph{Corresponding author: Y. Cai.})}
\thanks{Y. He, Y. Cai, H. Mao, and G. Yu are with the College of Information Science and Electronic Engineering, Zhejiang University, Hangzhou 310007, China, and also with Zhejiang Provincial Key Laboratory of Information Processing, Communication and Networking (IPCAN), Hangzhou 310007, China (e-mail: 2014hyh@zju.edu.cn;  ylcai@zju.edu.cn; hmao@zju.edu.cn; yuguanding@zju.edu.cn).}}

\maketitle
\vspace{-7ex}
\begin{abstract}
Integrated sensing and communication (ISAC) has been regarded as one of the most promising technologies for future wireless communications. However, the mutual interference in the communication radar coexistence system cannot be ignored. Inspired by the studies of reconfigurable intelligent surface (RIS), we propose a double-RIS-assisted coexistence system where two RISs are deployed for enhancing communication signals and suppressing mutual interference. We aim to jointly optimize the beamforming of RISs and radar to maximize communication performance while maintaining radar detection performance. The investigated problem is challenging, and thus we transform it into an equivalent but more tractable form by introducing auxiliary variables. Then, we propose a penalty dual decomposition (PDD)-based algorithm to solve the resultant problem. Moreover, we consider two special cases: the large radar transmit power scenario and the low radar transmit power scenario. For the former, we prove that the beamforming design is only determined by the communication channel and the corresponding optimal joint beamforming strategy can be obtained in closed-form. For the latter, we minimize the mutual interference via the block coordinate descent (BCD) method. By combining the solutions of these two cases, a low-complexity algorithm is also developed. Finally, simulation results show that both the PDD-based and low-complexity algorithms outperform benchmark algorithms.
\end{abstract}

\section{Introduction}
The future wireless communication system will develop towards high data rate and intelligentization to provide better services \cite{6G}. To enable high data rate transmission, there are two mainstream solutions. One is to push the wireless communication spectrum towards the higher frequency, such as millimeter waves \cite{mmwave} and even Terahertz waves \cite{tz};  the other is to share frequency bandwidth with other systems. Meanwhile, to realize the intelligent wireless communication system, artificial intelligence (AI) is regarded as one of the most potential techniques \cite{AI1,AI2}, and massive sensing data are necessary for designing and training the AI network. Further, the data can be obtained via deploying radar systems in communication systems. Due to the above two demands, integrated sensing and communication (ISAC) \cite{ISAC1,ISAC2,ISAC3} has recently attracted widespread attention in both academia and industry since it can collect sensing data and improve spectrum efficiency simultaneously. Specifically, the studies of ISAC focus on two aspects: 1) the dual-functional radar and communication (DFRC) system \cite{DF1,DF2,DF3}, in which radar and communication devices share the same hardware; 2) the communication radar coexistence system \cite{coexistence1,coexistence2}, in which radar and communication devices are separated and share the same frequency bandwidth. In this paper, we focus on the latter, and existing relevant works mainly aim at suppressing the mutual interference between two systems via joint radar and communication beamforming design.

However, communication devices still suffer from performance degradation in the communication radar coexistence system. To further mitigate the mutual interference, we introduce the reconfigurable intelligent surface (RIS) technique \cite{RIS1}, consisting of a large number of reconfigurable passive reflection elements, into the coexistence system.\footnote{RIS has been widely investigated in the communication system, and existing studies show that it improves the communication performance at a low cost via optimizing the beamforming matrix of RIS \cite{RIS2, RIS-channel,RIS3,RIS4,RIS5}. Specially, it plays an important role against the spectrum sharing issue in the cognitive radio system \cite{RIS6,RIS7}. Therefore, we are motivated to employ RIS in the communication radar coexistence system.} By deploying two RISs at the communication transmitter and receiver, respectively, the signal propagation among the transmitter, receiver, and radar can be reconfigured. It brings two benefits: enhance communication signal and suppress the mutual interference, and thus  can improve the performance of the coexistence system.

\subsection{Related Work}
Recently, the null-space projection \cite{NSP1,NSP2}, sophisticated optimization techniques via joint beamforming design \cite{joint1,joint2}, and subcarrier (spectrum) allocation \cite{allocation1}  have been widely investigated in the existing works for addressing the interference between radar and communication systems.  A few works have introduced RIS into ISAC \cite{RIS-a1,RIS-a2,RIS-a3,RIS-a4}.  The work in \cite{RIS-a1} considered a RIS-assisted multiple-input multiple-output (MIMO)  DFRC system where a RIS is deployed near a communication device to reduce mutual interference. To achieve it, an alternating optimization (AO)-based method was developed.  In \cite{RIS-a3}, the RIS has been employed for localization and communication when there is no direct path between the DFRC base station (BS) and sensing target. The RIS can be adaptively partitioned into two parts for communication and localization, respectively. The authors proposed a RIS passive beamforming algorithm and a corresponding target localization algorithm. The authors of \cite{RIS-a2} employed one  RIS for both sensing and communication in a more general scenario where the direct path between the DFRC BS and target exists. The goal is to maximize the radar signal-to-interference-plus-noise ratio (SINR)  under the communication SINR constraint, and the AO approach was utilized to solve the corresponding problem. Different from the above works,  \cite{RIS-a4} focused on the coexistence system and studied the spectrum sharing problem between MIMO radar and multi-user communication systems with the aid of RIS. The radar detection probability can be maximized  via beamforming optimization. {We should note that the previous studies only consider one RIS and the deployment of double distributed RISs has not been studied in the communication radar coexistence system. Meanwhile, it has been validated that deploying two RISs in communication systems can bring higher performance improvement compared to the single-RIS-assisted system \cite{RIS3}. Inspired by this, we seek to investigate a double-RIS-assisted communication radar coexistence system to further  mitigate the mutual interference by efficient joint beamforming design.}

\subsection{Main Contributions}
In this paper, we consider a classic scenario where a pair of communication transmitter and receiver share the same frequency bandwidth with a radar.\footnote{Our proposed techniques can also be extended to the general case with multiple receivers.} Unlike the single-RIS-assisted system in \cite{RIS-a4} and the conventional system without RIS, we propose a double-RIS-assisted communication radar coexistence system where two RISs are placed near the communication transmitter and receiver, respectively. {Particularly, the RIS placed near the transmitter is used to suppress interference from the transmitter to the radar, and the RIS placed near the receiver is used to cancel interference from the radar to the  receiver.} We aim at maximizing the communication performance by jointly optimizing the active beamforming matrix at the radar and the passive beamforming matrices at the RISs while ensuring the radar detection performance under the radar transmit power constraint. To solve the formulated problem, we propose a double-loop penalty dual decomposition (PDD)-based algorithm \cite{PDD,PDD1,PDD2}. Specifically, in the inner loop, the concave-convex procedure (CCCP) \cite{CCCP} is adopted for dealing with the difference-of-convex (DC) function, and the variables are updated in a block coordinate descent (BCD) fashion. In the outer loop, the Lagrange multipliers or the penalty parameter are updated. The proposed PDD-based algorithm converges to a stationary point of the original problem. After that, we study two special cases and develop a low-complexity algorithm. The main contributions are summarized as follows.
\begin{itemize}
	\item To suppress the mutual interference, we propose a novel double-RIS-assisted communication radar coexistence system with two RISs equipped near the transmitter and receiver, respectively. We then seek to maximize the communication SINR by jointly optimizing the active and passive beamforming matrices under the radar detection constraint.
	\item By introducing auxiliary variables, we transform the original problem into a more tractable form and develop a  PDD-based algorithm to solve it, which can be guaranteed to converge to a stationary point of the original problem. Moreover, the corresponding computational complexity of the proposed algorithm is also analyzed. 
	\item We consider two special cases: the large radar transmit power scenario and the low radar transmit power scenario. For the former, we derive the closed-form optimal solution of the joint beamforming design. For the latter, the mutual interference is minimized via the BCD method. Then, a low-complexity algorithm is developed by combining these two cases. 
	\item Simulation results are presented to validate our analysis and verify the effectiveness of the proposed double-RIS-assisted system by comparing it with the conventional systems. Besides, The performance comparison among the proposed algorithms and benchmark algorithms is provided.
\end{itemize}

\subsection{Organization}
The rest of this paper is organized as follows. Section II introduces the system model and formulates the optimization problem. In Section III, we develop a PDD-based algorithm to solve the problem. A low-complexity algorithm is developed in Section IV. The simulation results are presented in  Section V, and the whole paper is concluded in Section VI.

\emph{Notations:} In this paper, scalars are denoted by lower case, vectors are denoted by boldface lower case, and matrices are denoted by boldface upper case. $\bm{I}$ represents an identity matrix and $\bm{0}$ denotes an all-zero vector. $(\cdot)^*$, $(\cdot)^T$, and $(\cdot)^H$ denote complex conjugate, transpose, and Hermitian transpose, respectively. For a matrix $\bm{A}$, $\text{diag}(\bm{A})$ denotes a vector whose elements are the corresponding ones on the main diagonal of $\bm{A}$, $\bm{A}(:,n)$ denotes the $n$-th column vector,  $\bm{A}(n,:)$ denotes the $n$-th row vector, and $||\bm{A}||$ denotes its Frobenius norm. For a vector $\bm{a}$, $\text{Diag}(\bm{a})$ denotes a diagonal matrix with each diagonal element being the corresponding element in $\bm{a}$, $\bm{a}(n)$ denotes the $n$-th element, and $||\bm{a}||$ represents its Euclidean norm. $\Re\left(\cdot\right)$ denotes the real value of a complex scalar and $|\cdot|$ represents the absolute value of a complex scalar. $\mathbb{C}^{m\times n}$ ($\mathbb{R}^{m\times n}$) denotes the space of $m\times n$ complex (real) matrix. 
\begin{figure}[htbp]
	\centering
	\includegraphics[width=0.75\linewidth]{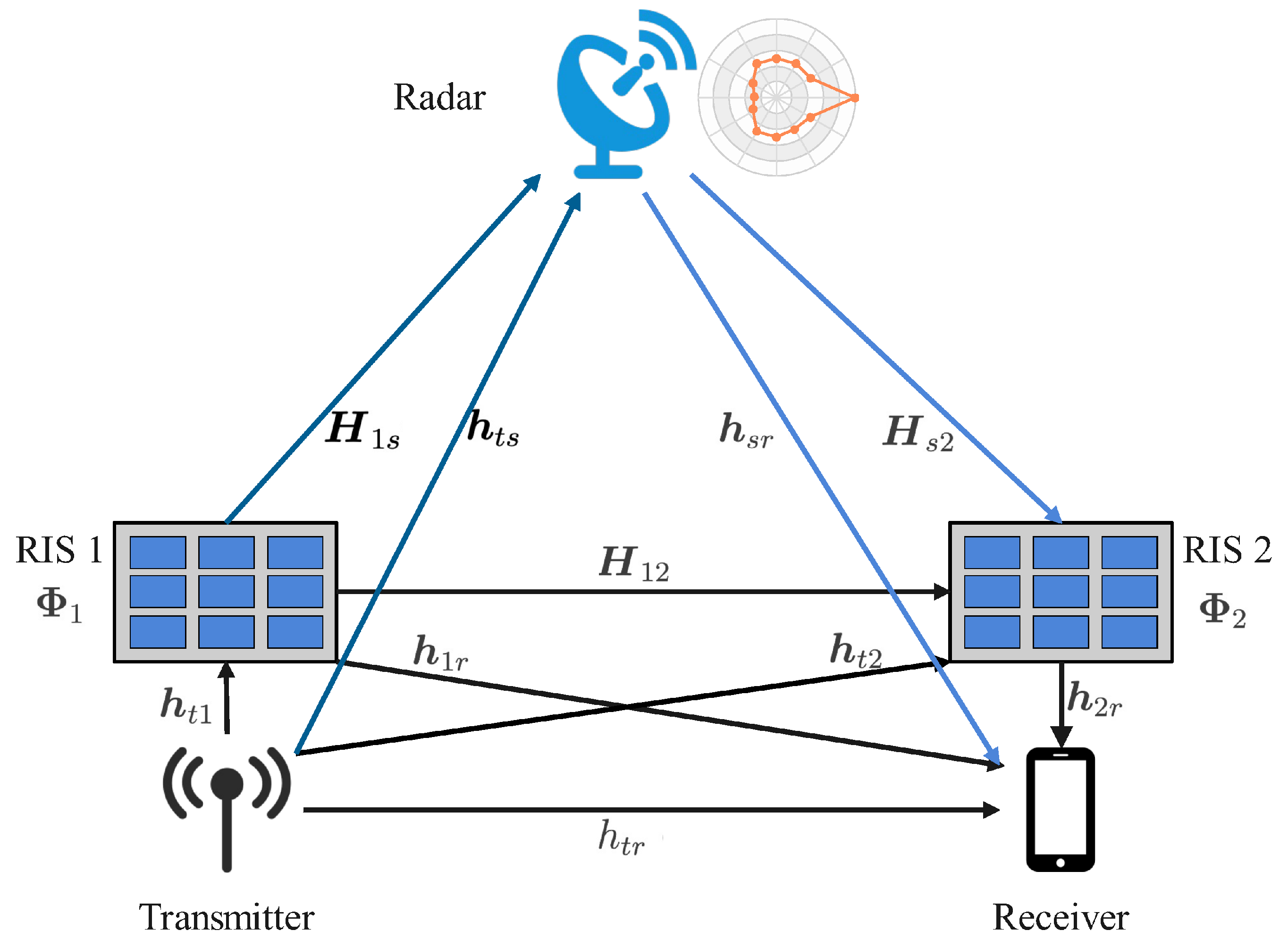}
	\caption{Double-RIS-assisted communication radar coexistence system.}
	\label{fig:sys}
	\vspace{-2ex}
\end{figure}

    \begin{table*}[t]
      \centering
      \caption{Notations of wireless channels} \label{tab:link}
      \begin{tabular}{|l|l||l|l||l|l|}
        \hline   
        Wireless channel & Notation  & Wireless channel & Notation & Wireless channel & Notation 
        \\ \hline \hline  
        Transmitter--Receiver & $h_{tr}\in\mathbb{C}$  &  RIS 2--Receiver & $\bm{h}_{2r}\in\mathbb{C}^{N_2\times 1}$ & RIS 1--Receiver & $\bm{h}_{1r}\in\mathbb{C}^{N_1\times 1}$
        \\ \hline 
        Transmitter--RIS 1 & $\bm{h}_{t1}\in\mathbb{C}^{N_1\times 1}$  & Radar--Receiver
         &$\bm{h}_{sr}\in\mathbb{C}^{M\times 1}$ & Radar--RIS 2 & $\bm{H}_{s2}\in\mathbb{C}^{N_2\times  M}$ 
        \\ \hline
       Transmitter--RIS 2 & $\bm{h}_{t2}\in\mathbb{C}^{N_2\times 1}$ & RIS 1--Radar & $\bm{H}_{1s}\in\mathbb{C}^{M\times N_1}$ & \emptyDiag{Wireless channel} &\emptyDiag{$\bm{H}_{s2}\in\mathbb{C}^{M\times N_2}$}
        \\ \hline 
        Transmitter--Radar
         &$\bm{h}_{ts}\in\mathbb{C}^{M\times 1}$  & RIS 1--RIS 2 & $\bm{H}_{12}\in\mathbb{C}^{N_2\times N_1}$ & \emptyDiag{Wireless channel}&\emptyDiag{$\bm{H}_{s2}\in\mathbb{C}^{M\times N_1}$}
        \\ \hline
      \end{tabular}
      \vspace{-2ex}
    \end{table*}

\section{System Model and Problem Formulation}
In this section, we first introduce the model of the double-RIS-assisted communication radar coexistence system and then mathematically formulate the optimization problem of interest.
\subsection{System Model}

In this paper, we consider a double-RIS-assisted communication radar coexistence system containing a phased-array radar and a pair of communication transmitter and receiver, as shown in Fig. \ref{fig:sys}. The radar and communication devices share the same frequency band. The radar is equipped with $M$ transmit and receive antennas. It aims at successively detecting the targets in $K$  directions in a detection epoch. For simplicity, we assume that the communication transmitter and receiver are both equipped with a single antenna. Two RISs, namely RIS 1 and RIS 2, are placed near the transmitter and receiver, respectively, to reconfigure wireless channels in order to enhance the communication signal while suppressing the mutual interference between radar and communication systems. Specifically, they are equipped with $N_1$ and $N_2$ reflecting elements, respectively.  The wireless channels in the proposed system are listed in Table \ref{tab:link}.\footnote{To investigate the joint beamforming design, we assume that the radar, communication devices, and RISs are connected to a central controller. The controller can effectively schedule and obtain all the required channel information based on efficient RIS-related channel estimation algorithms \cite{channel_est1,channel_est2}.}

\subsection{Radar Model}
Within a radar detection epoch, there are $K$ detection directions, denoted by $\{\theta_k\}, \forall k \in \mathcal{K}\triangleq \{1,\ldots, K\}.$ The length of detection time in each direction $\theta_k$, i.e., pulse repetition interval (PRI), is denoted by $L$. Thus, the duration of one detection epoch is $KL$. In each PRI, the probing pulse is transmitted in the first time index, i.e., $l=1$, and the echo signal from the target is assumed to be received at the time index $l_r$. Then, the probing pulse of the radar can be expressed as  
\begin{equation} \label{eq:radar_ts}
\bm{x}^r[l] = \left\{
\begin{array}{ll}
\bm{u}_ks^r_k, & l = (k-1)L+1,\forall k,\\
\bm{0},	& l\neq  (k-1)L+1,\forall k,
\end{array}\right.
\end{equation}
where $\bm{u}_k\in\mathbb{C}^{M\times 1}$ and $s_k^r$ denote the radar transmit beamforming vector and the radar signal for direction $\theta_k$, respectively. Therefore, the transmit power consumption for $\theta_k$ is $||\bm{u}_k||^2$ since we assume $\mathbb{E}\left\{\left(s_k^r\right)^*s_k^r\right\}=1$.

Then, the echo from the target in direction $\theta_k$ is given by 
\begin{equation}
    \bm{y}^{r,e}_k[l] =  \alpha_k \bm{a}(\theta_k)\bm{a}^T(\theta_k)\bm{x}^r[l-l_r],
\end{equation}
where $\bm{a}(\theta) \triangleq \left[1,e^{j\frac{2\pi\Delta}{\lambda_o}\pi\sin(\theta)},\cdots,e^{j\frac{2\pi\Delta}{\lambda_o} (M-1)\sin(\theta)} \right]^T \in \mathbb{C}^{M\times 1} $ with $\Delta$ being the antenna spacing and $\lambda_o$ being the wavelength, and $\alpha_k \bm{a}(\theta_k)\bm{a}^T(\theta_k)\in\mathbb{C}^{M\times M}$ represents the channel matrix. In addition, the received signal at the radar contains the interference from the communication transmitter. There are two links between the transmitter and the  radar, namely, the ``transmitter--radar'' link and the ``transmitter--RIS 1--radar'' link.\footnote{Based on the product-distance path loss model \cite{RIS-channel}, the average received signal power through the ``transmitter--RIS 2--radar'' link is much lower than that through the ``transmitter--radar'' link or the ``transmitter--RIS 1--radar'' link. Therefore, we neglect the ``transmitter--RIS 2--radar'' link.}
Thus, the interference from the transmitter to the radar can be expressed as 
\begin{equation}
\bm{y}^{r,i}_k[l] =\left(\bm{h}_{ts}+\bm{H}_{1s}\bm{\Phi}_{1}\bm{h}_{t1}\right)\sqrt{p^c}s^c[l],
\end{equation}
where $\bm{\Phi}_1=\text{Diag}(e^{j\phi_{1,1}},\cdots,e^{j\phi_{1,N_1}})\in\mathbb{C}^{N_1\times N_1}$ denotes the diagonal passive beamforming matrix at RIS 1 with $0\le \phi_{1,n}\le 2\pi, \forall n$, $s^c[l]$ is the communication transmit signal, and  $p^c$ is the corresponding  transmit power. By defining $\bm{H}_{ts}\triangleq\bm{H}_{1s}\text{Diag}(\bm{h}_{t1})$ and $\bm{\phi}_1 \triangleq \left[e^{j\phi_{1,1}},\cdots,e^{j\phi_{1,N_1}}\right]^T$, the interference can be rewritten as $\bm{y}^{r,i}_k[l] = \left(\bm{h}_{ts}+\bm{H}_{ts}\bm{\phi}_1\right)\sqrt{p^c}s^c[l]$.

Based on the above, for the target in direction $\theta_k$, the received signal at the radar can be expressed as 
\begin{align}
	\bm{y}^r_k[l] =& \alpha_k \bm{a}(\theta_k)\bm{a}^T(\theta_k)\bm{x}^r[l-l_r] \nonumber \\
	&+ \left(\bm{h}_{ts}+\bm{H}_{ts}\bm{\phi}_1\right)\sqrt{p^c}s^c[l]+\bm{n}_0[l],
\end{align}
where $\bm{n}_0[l]\sim \mathcal{CN}\left(0,\sigma^2\bm{I}\right)$ is the complex circular Gaussian noise vector with zero mean and covariance $\sigma_r^2\bm{I}$, including the clutter and the additive white Gaussian noise.

After receiving the signal, the receive  beamforming vector $\bm{w}_k$ is utilized for detecting the echo from direction $\theta_k$ and the corresponding SINR can be given as
\begin{equation}
	\text{SINR}^r_k =   \dfrac{\left|\bm{w}_k\alpha_k\bm{a}(\theta_k)\bm{a}^T(\theta_k)\bm{u}_k\right|^2}{\sigma^2\left|\bm{w}_k\right|^2+p^c\left|\bm{w}_k\left(\bm{h}_{ts}+\bm{H}_{ts}\bm{\phi}_1\right)\right|^2}, ~\forall k.
\end{equation}

\subsection{Communication Model}
With the aid of RISs, the  transmitter communicates with the receiver through four wireless links, namely, the ``transmitter--receiver'' link, the ``transmitter--RIS 1--receiver'' link, the ``transmitter--RIS 2--receiver'' link, and the ``transmitter--RIS 1--RIS 2--receiver'' link.  In the meanwhile, the signal from radar, i.e., $\bm{x}^r[l]$, interferes with the communication receiver via two wireless links: the ``radar--receiver'' link and the ``radar--RIS 2--receiver'' link.  Therefore, the received signal at the receiver at  time index $l$ can be expressed as
\begin{align}
	&y^c[l] =\underbrace{\left(\bm{h}^H_{sr}+\bm{h}^H_{2r}\bm{\Phi}_2\bm{H}_{s2}\right)\bm{x}^r[l]}_{\text{interference from the radar}} + n_0[l]\nonumber\\
	&+\!\underbrace{\!\left(\!h_{tr}\!+\!\bm{h}^H_{1r}\bm{\Phi}_1\bm{h}_{t1}\!+\!\bm{h}^H_{2r}\bm{\Phi}_2\bm{h}_{t2}\!+\!\bm{h}_{2r}^H\bm{\Phi}_{2}\bm{H}_{12}\bm{\Phi}_1\bm{h}_{t1}\!\right)\!\sqrt{p^c}s^c[l]}_{\text{communication signal}},
\end{align}
where $\bm{\Phi}_2=\text{Diag}(e^{j\phi_{2,1}},\cdots,e^{j\phi_{2,N_2}})\in\mathbb{C}^{N_2\times N_2}$ denotes the diagonal passive beamforming matrix at RIS 2 with $0\le \phi_{2,n}\le 2\pi, \forall n$, and $n_0[l]\sim \mathcal{CN}\left(0,\sigma^2\right)$ represents the complex white Gaussian noise at the receiver with zero mean and variance $\sigma^2$.

By defining $\bm{g}_{tr} \triangleq \text{Diag}(\bm{h}^*_{t1})\bm{h}_{1r}$, $\bm{f}_{tr}\triangleq \text{Diag}(\bm{h}^*_{t2})\bm{h}_{2r}$,       $\bm{H}_{tr} \triangleq \text{Diag}(\bm{h}_{2r}^*)\bm{H}_{12} \text{Diag}(\bm{h}_{t1})$,   $\bm{H}_{sr}\triangleq\text{Diag}(\bm{h}^*_{2r})\bm{H}_{s2}$, and $\bm{\phi}_2 \triangleq \left[e^{j\phi_{2,1}},\cdots,e^{j\phi_{2,N_2}}\right]^T$, the  received signal can be rewritten as 
\begin{align}
 	y^c[l] =& \left(h_{tr}+\bm{g}_{tr}^H\bm{\phi}_1+\bm{f}_{tr}^H\bm{\phi}_2+\bm{\phi}_1^T\bm{H}_{tr}\bm{\phi}_2\right)\sqrt{p^c}s^c[l]\nonumber \\
 	&+\left(\bm{h}^H_{sr}+\bm{\phi}_2^T\bm{H}_{sr}\right)\bm{x}^r[l]+n_0[l].
\end{align}
Then, the overall energy of the received signal during  one radar detection epoch is given by
\begin{align}
	&\mathbb{E} \left(\sum_{l=1}^{KL} \left(y^c[l]\right)^H y^c[l] \right)\nonumber\\
	=&\underbrace{KLp^c\left|h_{tr}+\bm{g}_{tr}^H\bm{\phi}_1+\bm{f}_{tr}^H\bm{\phi}_2+\bm{\phi}_1^T\bm{H}_{tr}\bm{\phi}_2\right|^2}_{\text{communication signal}}\nonumber\\
	&+\underbrace{\sum_{k=1}^K\left|\bm{h}^H_{sr}\bm{u}_k+\bm{\phi}_2^T\bm{H}_{sr}\bm{u}_k\right|^2}_{\text{interference}}+\underbrace{KL\sigma^2}_{\text{noise}}.
\end{align}
We adopt the average communication SINR as the metric for the communication performance, which can be expressed as 
\begin{equation}
	\text{SINR}^c = \dfrac{KLp^c\left|h_{tr}+\bm{g}_{tr}^H\bm{\phi}_1+\bm{f}_{tr}^H\bm{\phi}_2+\bm{\phi}_1^T\bm{H}_{tr}\bm{\phi}_2\right|^2}{KL\sigma^2+\sum_{k=1}^K\left|\bm{h}^H_{sr}\bm{u}_k+\bm{\phi}_2^T\bm{H}_{sr}\bm{u}_k\right|^2}.
\end{equation}

\subsection{Problem Formulation}
In this paper, we aim at maximizing the communication performance   while guaranteeing the radar detection performance for each detection direction. The radar performance is positively associated with the radar SINR \cite{radar_performance}. Therefore, the communication SINR maximization problem can be formulated as 
\begin{subequations}\label{pb:o}
\begin{eqnarray}
\!\!\!\!\!\!\!\!	&\!\!\!\! \max\limits_{\left\{\substack{ \bm{u}_k, \bm{w}_k, \\\bm{\phi}_1, \bm{\phi}_2}\right\}}\!\!\!\! & \!\!\dfrac{KLp^c\left|h_{tr}\!+\!\bm{g}_{tr}^H\bm{\phi}_1\!+\!\bm{f}_{tr}^H\bm{\phi}_2\!+\!\bm{\phi}_1^T\bm{H}_{tr}\bm{\phi}_2\right|^2}{KL\sigma^2+\sum\limits_{k=1}^K\left|\bm{h}^H_{sr}\bm{u}_k+\bm{\phi}_2^T\bm{H}_{sr}\bm{u}_k\right|^2},\\
\!\!\!\!\!\!\!\!&\!\!\!\!\text{s.t.}\!\!\!\!&\!\! \dfrac{\left|\bm{w}_k^H\alpha_k\bm{a}(\theta_k)\bm{a}^T(\theta_k)\bm{u}_k\right|^2}{\sigma^2\left|\bm{w}_k\right|^2\!+\!p^c\left|\bm{w}_k^H\left(\bm{h}_{ts}\!+\!\bm{H}_{ts}\bm{\phi}_1\right)\right|^2}\! \ge\! \gamma^r,\forall k,~~~~~~\label{con:o_radar_sinr}\\
\!\!\!\!\!\!\!\!&	&\!\!\sum_{k=1}^K\left|\left|\bm{u}_k\right|\right|^2 \le P_{\max},\label{con:o_radar_power}\\
\!\!\!\!\!\!\!\!&	& \!\!|\bm{\phi}_1(n)| = 1, |\bm{\phi}_2(n)| = 1,~\forall n,	\label{con:o_ris}	
\end{eqnarray}
\end{subequations}
where $\gamma^r$ denotes the SINR threshold and constraint (\ref{con:o_radar_sinr}) guarantees the radar detection performance for each direction. Constraint  (\ref{con:o_radar_power}) denotes the radar transmit power limitation, where $P_{\max}$ is the budget of the total radar power consumption. Constraint (\ref{con:o_ris}) denotes the uni-modulus constraint on all elements of the RIS passive beamforming vector.

Problem (\ref{pb:o}) is difficult to solve due to the highly coupled and nonconvex objective function and constraints.  Thus, in the following section, we seek to propose a joint beamforming design algorithm to solve this problem.

\section{Joint Beamforming Design Algorithm}

In this section, we first provide a brief overview of the PDD framework, and then transform problem  (\ref{pb:o}) into a more tractable but equivalent form by introducing some auxiliary variables and equality constraints. After that, we propose a novel PDD-based algorithm to  solve the converted problem, where  the augmented Lagrangian (AL) terms are combined into the objective function and  an AL problem is formulated. The proposed PDD-based algorithm has double loops. In the outer loop, we update the dual variables or the penalty parameter, while in the inner loop, we solve the AL problem.

{\vspace{-1.5ex}
	\begin{algorithm}[b]
	\caption{PDD Framework for Problem $\mathcal{P}$.}
	\label{ALG2}
	\DontPrintSemicolon
	Initialize $\bm{x}^{(0)}\in\mathcal{X}$, $\rho^{(0)}>0$, $\eta^{(0)}>0$, $\bm{\lambda}$, and set $0\le c\le 1$, $i=0$.\;
	\Repeat{some termination criterion is met.}{
		\% Solve the AL problem \; 
		$\bm{x}^{(i+1)}$ = optimize$\left(\mathcal{P} \left( \rho^{(i)},\bm{\lambda}^{(i)} \right), \bm{x}^{(i)} \right) $; \;
		\eIf{$||\bm{h}(\bm{x}^{(i+1)})||_{\infty} \le \eta^{(i)}$}{
			$\bm{\lambda}^{(i+1)} = \bm{\lambda}^{(i)} +\dfrac{1}{\rho^{(i)}} \bm{h}(\bm{x}^{(i+1)})$, 
			$\rho^{(i+1)} = \rho^{(i)}$;\;
		}
		{$\bm{\lambda}^{(i+1)} = \bm{\lambda}^{(i)} $, 
			$\rho^{(i+1)} = c\rho^{(i)}$;\;
		}
		$\eta^{(i+1)} = 0.7||\bm{h}(\bm{x}^{(i)})||_{\infty}$;\;
		$i=i+1$;\;
	}
\end{algorithm}}

\subsection{PDD Framework}
We briefly introduce the general framework of the PDD method in the following. Consider the following problem
\begin{subequations}
\begin{eqnarray}
	\mathcal{P}: & \min\limits_{\bm{x}\in \mathcal{X}} &  f(\bm{x}),\\
	&\text{s.t.}& 	\bm{h}(\bm{x}) = 0,\\
	&	& \bm{g}(\bm{x}) \le 0,
\end{eqnarray}
\end{subequations}
where $f(\bm{x})$ is a scalar continuously differentiable function, $\mathcal{X}\subseteq \mathbb{R}^{n}$ is a closed convex set, $\bm{h}(\bm{x})\in\mathbb{R}^{p\times1}$ is a vector of $p$ continuously differentiable functions, and $\bm{g}(\bm{x})\in\mathbb{R}^{q\times1}$ is a vector of $q$ continuously differentiable but possibly nonconvex functions. As shown in Algorithm  \ref{ALG2}, the double-loop PDD framework can be employed for solving the general problem $\mathcal{P}$. To be specific, in the inner loop, it focuses on solving the following AL problem $\mathcal{P}\left( \rho^{(i)},\bm{\lambda}^{(i)} \right)$, which is subsumed by the “optimize” function:
\begin{subequations}
\begin{eqnarray}
	\!\!\!\mathcal{P}\left( \rho^{(i)},\bm{\lambda}^{(i)} \right): \!\!\! & \min\limits_{\bm{x}\in \mathcal{X}} &\!\!\!  \mathcal{L}^{(i)}(\bm{x}) \triangleq f(\bm{x}) \nonumber\\
	&   &\!\!\! + \dfrac{1}{2\rho^{(i)}} \left|\left| \bm{h}(\bm{x})+\rho^{(i)} \bm{\lambda}^{(i)}\right|\right|^2,\\
	&\!\!\!\text{s.t.}&  \bm{g}(\bm{x}) \le 0,
\end{eqnarray}
\end{subequations}
where $\mathcal{L}^{(i)}(\bm{x}) $ denotes the AL function with the dual variable $\bm{\lambda}^{(i)}$ and penalty factor $\rho^{(i)}$, and $i$ denotes the current iteration number of the outer loop. According to \cite{PDD}, solving problem $\mathcal{P}\left( \rho^{(i)},\bm{\lambda}^{(i)} \right)$ produces an identical solution to problem $\mathcal{P}$ when $\rho^{(i)}\rightarrow 0$.  In the outer loop,  the  dual variable $\bm{\lambda}^{(i)}$  or  penalty factor $\rho^{(i)}$ are updated in terms of the constraint violation $||\bm{h}(\bm{x}^{(i)})||_{\infty}$ that is used for measuring the violation of the constraint $\bm{h}(\bm{x})=0$. The convergence of the PDD method has been  proved in  \cite{PDD1} and \cite{PDD2},   demonstrating that $\bm{x}^{(i)}$ obtained by the PDD method converges to a stationary point of problem $\mathcal{P}$. Moreover, in Algorithm \ref{ALG2}, $\rho$, $c$, and $\eta$ are set empirically.

\subsection{Problem Transformation and AL Problem}
Before applying the PDD framework, we can find that  there are two difficulties for solving the  communication SINR maximization problem  (\ref{pb:o}), as follows.
\begin{itemize}
\item The objective function contains a fractional coupling term.
\item Constraint (\ref{con:o_radar_sinr}) is highly coupled and the PDD framework cannot be applied directly.
\end{itemize}
To deal with them, in the following, we introduce some  auxiliary variables and transform  problem (\ref{pb:o}) into an equivalent but more tractable form. First of all, regarding the objective function of problem (\ref{pb:o}), we have the following Lemma. 
\begin{lem} \label{lem:v}
By introducing an auxiliary variable $v\in\mathbb{C}$, the objective function of problem (\ref{pb:o}) can be equivalently reformulated as the following one without influencing the optimality:
\begin{eqnarray}
	\!\!\!\!\!\!\!\!\!\!&\min\limits_{\left\{\substack{ \bm{u}_k, \bm{w}_k, \\v, \bm{\phi}_1, \bm{\phi}_2}\right\}}& |v|^2\left({KL\sigma^2+\sum_{k=1}^K\left|\bm{h}^H_{sr}\bm{u}_k+\bm{\phi}_2^T\bm{H}_{sr}\bm{u}_k\right|^2} \right)\nonumber\\
	\!\!\!\!\!\!\!\!\!\!& &-\!2\Re\left(\!v^*\! \sqrt{KLp^c}\left(h_{tr}+\bm{g}_{tr}^H\bm{\phi}_1 \right.\right.\nonumber\\
	\!\!\!\!\!\!\!\!\!\!& &~~~~~~~~~~~~~~~~~~~~\left.\left.+\bm{f}_{tr}^H\bm{\phi}_2\!+\!\bm{\phi}_1^T\bm{H}_{tr}\bm{\phi}_2\right)\right).
\end{eqnarray}
\begin{IEEEproof}
The proof can be found in \cite{FP} and  the detailed derivation is omitted for brevity.
\end{IEEEproof}
\end{lem}

Then, to apply the PDD framework, we need to deal with the highly coupling inequality constraint (\ref{con:o_radar_sinr}). Therefore, we introduce auxiliary variables $x_k$ and $y_k$, $\forall k$ with the following equality constraints
\begin{align}
x_k &= \bm{w}_k^H\bm{a}(\theta_k)\bm{a}^T(\theta_k)\bm{u}_k ,~\forall k, \label{con:x}\\
y_k&=\bm{w}_k^H\left(\bm{h}_{ts}+\bm{H}_{ts}\bm{\phi}_1\right),~\forall k.\label{con:y}
\end{align}
Thus, constraint (\ref{con:o_radar_sinr}) can be rewritten as 
\begin{equation}
	\gamma^r\left({\sigma^2\left|\bm{w}_k\right|^2+ p^c\left|y_k \right|^2}\right) - |\alpha_k|^2 {\left|x_k\right|^2} \le 0,~\forall k. \label{con:radar_sin}
\end{equation}
By dualizing and penalizing the constraints (\ref{con:x}) and (\ref{con:y})  into objective function with dual variables $\{\lambda_{k,1},\lambda_{k,2},\forall k\}$ and penalty factor $\rho$, we can obtain the following AL problem for the inner loop of the PDD framework
\begin{eqnarray}
	\!\!\!& \min\limits_{\mathbb{X}} &  |v|^2\left({KL\sigma^2+\sum_{k=1}^K\left|\bm{h}^H_{sr}\bm{u}_k+\bm{\phi}_2^T\bm{H}_{sr}\bm{u}_k\right|^2} \right) \nonumber \\
	\!\!\!&	&-2\Re\left(v^* \sqrt{KLp^c}\left(h_{tr}+\bm{g}_{tr}^H\bm{\phi}_1+\bm{f}_{tr}^H\bm{\phi}_2\right.\right.\nonumber\\
	\!\!\!& &~~~~~~~~~~~~~~~~~~~~~~~~~\left.\left.+\bm{f}_{tr}^H\bm{\phi}_2\!+\!\bm{\phi}_1^T\bm{H}_{tr}\bm{\phi}_2\right)\right)\nonumber \\
	\!\!\!&	&+ \dfrac{1}{2\rho}\sum_{k=1}^K\left(|x_k - \bm{w}_k^H\bm{a}(\theta_k)\bm{a}^T(\theta_k)\bm{u}_k+\rho\lambda_{k,1}|^2\right)\nonumber\\
	\!\!\!&	&+\dfrac{1}{2\rho}\sum_{k=1}^K\left(|y_k- \bm{w}_k^H\left(\bm{h}_{ts}+\bm{H}_{ts}\bm{\phi}_1\right)+\rho\lambda_{k,2}|^2\right),~~~~\label{pb:AL}\\
	\!\!\!&\text{s.t.}&  \text{ (\ref{con:o_radar_power}), (\ref{con:o_ris}), and (\ref{con:radar_sin})}, \nonumber
\end{eqnarray}
where $\mathbb{X}\triangleq \{v, \bm{u}_k, \bm{w}_k, \bm{\phi}_1, \bm{\phi}_2,x_k,y_k\}$ represents the set of optimization variables. 

With the aid of the PDD framework and Lemma \ref{lem:v},  we can see that problem (\ref{pb:AL}) is equivalent to problem (\ref{pb:o}) when $\rho \rightarrow 0$. In the following subsection, we will focus on solving  the AL problem (\ref{pb:AL}) in the inner loop.

\subsection{Proposed CCCP-Based Algorithm for Solving Problem  (\ref{pb:AL})}
Let us solve the AL problem  (\ref{pb:AL}) in the inner loop. It is still difficult since constraint (\ref{con:radar_sin}) is nonconvex, which is a  DC function. Fortunately, according to the CCCP approach \cite{CCCP}, in the inner loop, we can approximate this constraint by linearization. First, constraint (\ref{con:radar_sin}) can be rewritten as 
\begin{equation}
    f_{1,k}(\bm{w}_k, y_k) - f_{2,k}(x_k) \le 0,~ \forall k,
\end{equation}
where 
\begin{align}
    &f_{1,k}(\bm{w}_k, y_k) = \gamma^r\left({\sigma^2\left|\bm{w}_k\right|^2+ p^c\left|y_k \right|^2}\right),\\
    &f_{2,k}(x_k) = |\alpha_k|^2 {\left|x_k\right|^2}.
\end{align}
Then, we approximate the convex function $f_{2,k}(x_k)$ in the $j$-th iteration around the current point $x_{k}^{(j)}$ by utilizing its first order Taylor expansion, as 
\begin{equation}
	\hat{f}_{2,k}\left(x_{k}^{(j)},x_k\right)= |\alpha_k|^2 \left( 2\Re\left(\left(x_k^{(j)}\right)^* x_k \right)-\left|x_k^{(j)}\right|^2 \!\right).
\end{equation}
Then, constraint (\ref{con:radar_sin}) can be approximated as
\begin{equation}
	 f_{1,k}(\bm{w}_k, y_k) - \hat{f}_{2,k}\left(x_{k}^{(j)},x_k\right)\le 0,~\forall k. \label{con:radar_sin3}
\end{equation}
The solution obtained based on the constraint (\ref{con:radar_sin3}) is feasible to  problem (\ref{pb:AL}) according to the CCCP approach. The corresponding problem is 
\begin{eqnarray}
	\!\!\!& \min\limits_{\mathbb{X}} &  |v|^2\left({KL\sigma^2+\sum_{k=1}^K\left|\bm{h}^H_{sr}\bm{u}_k+\bm{\phi}_2^T\bm{H}_{sr}\bm{u}_k\right|^2} \right) \nonumber \\
	\!\!\!&	&-2\Re\left(v^* \sqrt{KLp^c}\left(h_{tr}+\bm{g}_{tr}^H\bm{\phi}_1+\bm{f}_{tr}^H\bm{\phi}_2\right.\right.\nonumber\\
	\!\!\!& &~~~~~~~~~~~~~~~~~~~~~~~~~\left.\left.+\bm{f}_{tr}^H\bm{\phi}_2\!+\!\bm{\phi}_1^T\bm{H}_{tr}\bm{\phi}_2\right)\right)\nonumber \\
	\!\!\!&	&+ \dfrac{1}{2\rho}\sum_{k=1}^K\left(|x_k - \bm{w}_k^H\bm{a}(\theta_k)\bm{a}^T(\theta_k)\bm{u}_k+\rho\lambda_{k,1}|^2\right)\nonumber\\
	\!\!\!&	&+\dfrac{1}{2\rho}\sum_{k=1}^K\left(|y_k- \bm{w}_k^H\left(\bm{h}_{ts}+\bm{H}_{ts}\bm{\phi}_1\right)+\rho\lambda_{k,2}|^2\right),~~~~\label{pb:CCCP}\\
	\!\!\!&\text{s.t.}& \text{ (\ref{con:o_radar_power}), (\ref{con:o_ris}), and  (\ref{con:radar_sin3})}, \nonumber
\end{eqnarray}
which becomes a convex problem. By utilizing the CCCP approach, $x_{k}^{(j)}$ converges to a stationary point of problem (\ref{pb:CCCP}). 

To solve problem (\ref{pb:CCCP}), we can employ the BCD method \cite{PDD}. The variables can be divided into four blocks and they are updated successively by solving the corresponding subproblems. Specifically, the details of the subproblems and their solutions are shown as follows.

In \textbf{Step 1}, we optimize $v$, $\{\bm{w}_k,  x_k,y_k\}$ in parallel by fixing the other variables. For optimizing  $v$, we consider the following subproblem:
\begin{eqnarray}
	& \min\limits_{v} &  |v|^2a -2\Re\left(v^*b\right),\label{pb:v}
\end{eqnarray}
where $a\!\triangleq \! {KL\sigma^2\!+\!\sum_{k=1}^K\left|\bm{h}^H_{sr}\bm{u}_k+\bm{\phi}_2^T\bm{H}_{sr}\bm{u}_k\right|^2} $ and $b\!\triangleq \! {\sqrt{KLp^c}\left(h_{tr}+\bm{g}_{tr}^H\bm{\phi}_1+\bm{f}_{tr}^H\bm{\phi}_2+\bm{\phi}_1^T\bm{H}_{tr}\bm{\phi}_2\right)}$. Based on the first order optimality condition, the optimal solution to problem (\ref{pb:v}) is given by
\begin{equation}
	v^{\star} = {b}/{a}.
\end{equation}

The subproblem for $\{\bm{w}_k,  x_k,y_k\}$  is given by 
\begin{eqnarray}
	\!\!\!\!& \min\limits_{\bm{w}_k,  x_k,y_k}\!\!\! &  \!\!\! |x_k \!-\! \bm{w}_k^H \bm{q}_k\!+\!\rho\lambda_{k,1}|^2 \!+\! |y_k\!-\!\bm{w}_k^H \bm{f} \!+\!\rho\lambda_{k,2}|^2,~~~~~~~\label{pb:step1}\\
	\!\!\!\!&\text{s.t.}\!\!\!&  \!\!\!\text{(\ref{con:radar_sin3})},  \nonumber 
\end{eqnarray}
where $\bm{q}_k \triangleq \bm{a}(\theta_k)\bm{a}^T(\theta_k)\bm{u}_k  $ and $\bm{f}  \triangleq \left(\bm{h}_{ts}+\bm{H}_{ts}\bm{\phi}_1\right)$. The optimal solution to this subproblem, denoted by $\{\bm{w}_k^{\star}$, $x_k^{\star}$, $ y_k^{\star}\}$, can also be solved by using the Lagrange multiplier method  and the first order optimality condition. The detailed derivation is provided in Appendix \ref{app:step}.

In \textbf{Step 2}, we optimize $\bm{u_k}$ by fixing the other variables and the corresponding subproblem is given by
\begin{subequations}\label{pb:u}
\begin{eqnarray} 
	\!\!\!\!& \min\limits_{\bm{u}_k} & \!\!\!\! |v|^2 \sum_{k=1}^K\left|\bm{p}^H\bm{u}_k\right|^2  \!+\! \dfrac{1}{2\rho}\sum_{k=1}^K|  \bm{h}_k^H \bm{u}_k\!-\!x_k \!-\!\rho\lambda_{k,1}|^2,~~~~~~\\
	\!\!\!\!&\text{s.t.}& \!\!\!\! \text{(\ref{con:o_radar_power})},
\end{eqnarray}
\end{subequations}
where $\bm{p} \triangleq (\bm{h}^H_{sr} +\bm{\phi}_2^T\bm{H}_{sr})^H$ and $\bm{h}_k \triangleq \left(\bm{w}_k^H\bm{a}(\theta_k)\bm{a}^T(\theta_k)\right) ^H$. 
The optimal solution to this subproblem, denoted by $\bm{u}_k^{\star}$, can be obtained by utilizing the Lagrange multiplier method and the first order optimality condition. The detailed derivation is also provided in Appendix \ref{app:step}.

In \textbf{Step 3}, we optimize $\bm{\phi}_1(n)$, $\forall n$ sequentially by fixing the other variables. The subproblem with respect to $\bm{\phi}_1(n)$ is given by
\begin{subequations}
\begin{eqnarray} 
	& \min\limits_{\bm{\phi}_1(n)}  &  -\Re\left(a_n \bm{\phi}_1(n) \right) +\dfrac{1}{\rho}\sum_{k=1}^K\left( \Re\left( b_{k,n}\bm{\phi}_1(n)\right)\right) \nonumber\\
	& &+\dfrac{1}{2\rho}\sum_{k=1}^K\left( | \bm{w}_k^H\bm{H}_{ts}(:,n) |^2 |\bm{\phi}_1(n) |^2\right) ,\\
	&\text{s.t}	& |\bm{\phi}_1(n)| = 1,
\end{eqnarray}
\end{subequations}
where 
$b_{k,n}\! \triangleq \! \left(\!\sum\limits_{i\neq n}\bm{w}_k^H\bm{H}_{ts}(:,i) \bm{\phi}_1(i)\!-\!(y_k\!-\!\bm{w}_k^H \bm{h}_{ts}\!+\!\rho\lambda_{k,2}) \right)^* $ $ \!\!\left(\bm{w}_k^H\bm{H}_{ts}(:,n)\right)$ and $a_n \!\triangleq\! 2v^* \!\sqrt{KLp^c} \left(\bm{g}_{tr}^H(n) \!+\! \bm{H}_{tr}(n,:)\bm{\phi}_2\right)$. Then, the optimal solution is 
\begin{equation} \label{res:phi_1}
	\bm{\phi}_1^{\star}(n) = \exp\left(j\pi -j \angle\left( \frac{\sum_{k=1}^K b_{k,n}}{\rho}-a_n \right)\right).   
\end{equation}

In \textbf{Step 4}, we optimize $\bm{\phi}_2(n)$, $\forall n$ sequentially by fixing the other variables. The corresponding subproblem is given by
\begin{subequations}
\begin{eqnarray}
	& \min\limits_{\bm{\phi}_2(n)} & 2|v|^2\sum_{k=1}^K\Re\left(c_{k,n}\bm{\phi}_2(n)\right)-2\Re\left(d_n\bm{\phi}_2(n)\right) \nonumber \\
	&   &+|v|^2\sum_{k=1}^K |\bm{H}_{sr}(n,:)\bm{u}_k|^2|\bm{\phi}_2(n)|^2 ,\\
	&\text{s.t}	& |\bm{\phi}_2(n)| = 1,
\end{eqnarray}
\end{subequations}
where $d_n  \triangleq  v^* \sqrt{KLp^c}\left(\bm{f}_{tr}^H(n)+ \bm{\phi}_1^T\bm{H}_{tr}(:,n)\right)$ and $c_{k,n}  \triangleq  \left( \sum\limits_{i\neq n}\bm{H}_{sr}(i,:)\bm{u}_k \bm{\phi}_2(i)\!+  \bm{h}^H_{sr}\bm{u}_k\!\right)^* \!\!\bm{H}_{sr}(n,:)\bm{u}_k$. Then, the optimal solution to this subproblem is given by
\begin{equation}\label{res:phi_2}
	\bm{\phi}_2^{\star}(n) = \exp\left(j\pi-j \angle \left( |v|^2\sum_{k=1}^K c_{k,n}-d_n\right)\right).
\end{equation}

So far, we have solved all subproblems with closed-form solutions and the  proposed CCCP-based algorithm for problem  (\ref{pb:AL}) is summarized  in Algorithm \ref{ALG1}, where we implement  four updating steps  in each iteration.	

{\vspace{-2ex}
	\begin{algorithm}[h]
		\caption{Proposed CCCP-Based Algorithm in the Inner Loop.}
		\label{ALG1}
		\DontPrintSemicolon
		Initialize variables $\mathbb{X}=\{v, \bm{u}_k, \bm{w}_k, \bm{\phi}_1, \bm{\phi}_2,x_k,y_k\}$. Set the tolerance of accuracy $\epsilon_1$, the maximum number of iterations $I_{\max}$, and the iteration number $j=0$.\;
		\Repeat{the gap between consecutive values of the
			objective function is under $\epsilon_1$ or $j>I_{\max}$.}{
			Update $v$, $\{\bm{w}_k,  x_k,y_k\}$ in \textbf{Step 1};\;
			Update $\bm{u}_k$ in \textbf{Step 2};\;
			Update $\bm{\phi}_1$ in \textbf{Step 3};\;
			Update $\bm{\phi}_2$ in \textbf{Step 4};\;
			$j=j+1$;\;
		}
	\end{algorithm}
	\vspace{-2ex}
}

\subsection{Summary of the Proposed PDD-based Algorithm}

\begin{figure}[t]
	\centering
	\includegraphics[width=0.7\linewidth]{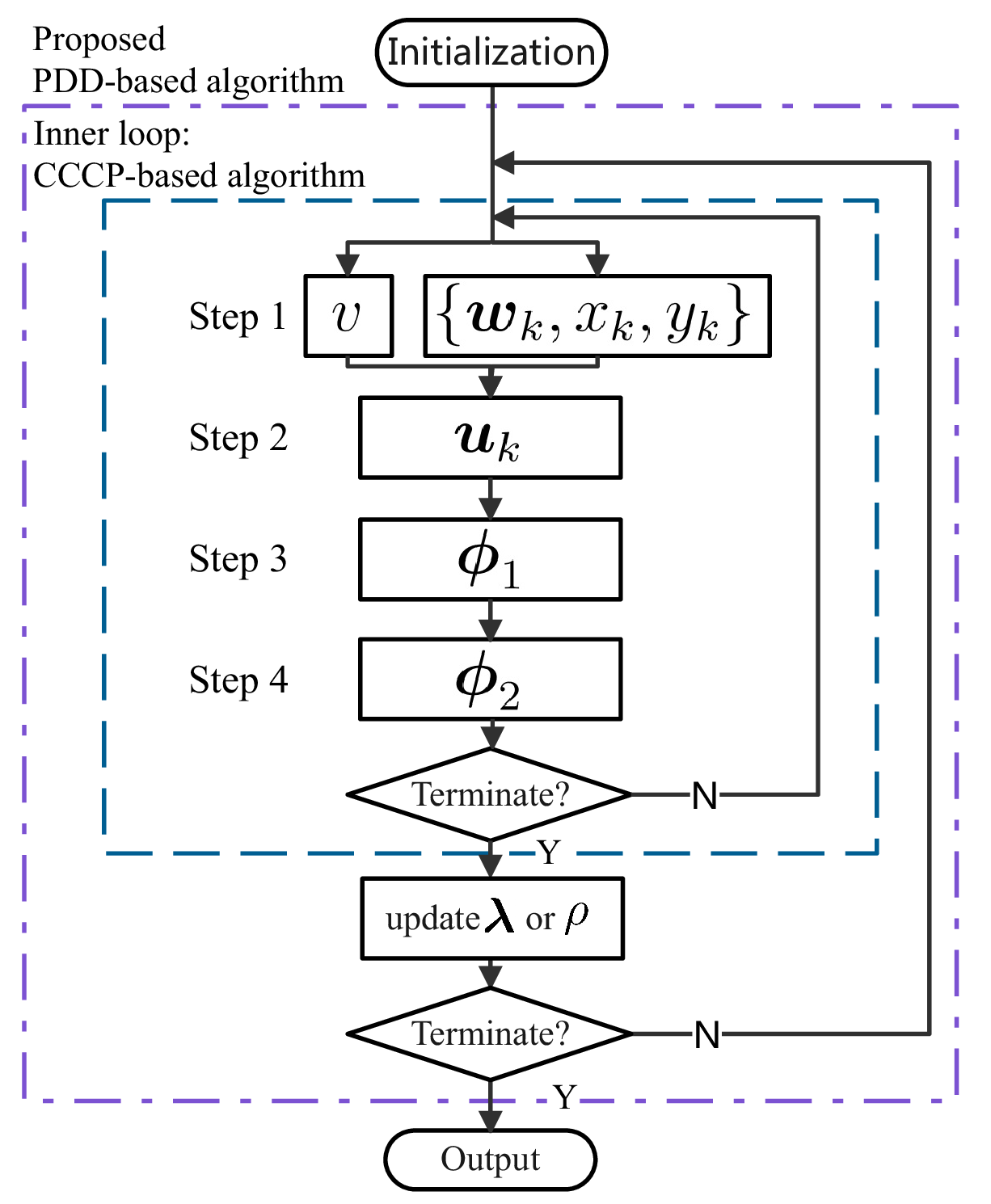}
	\vspace{-2ex}
	\caption{The structure of the proposed PDD-based algorithm.}
	\vspace{-2ex}
	\label{fig:flow}
\end{figure}

Recall that we aim at solving the original  problem (\ref{pb:o}) by utilizing the PDD framework as shown in Algorithm \ref{ALG2} and the structure of the proposed PDD-based algorithm is shown in Fig. \ref{fig:flow}. In the inner loop, we solve the AL problem  (\ref{pb:AL}) by utilizing the proposed CCCP-based algorithm, and then update the  dual variables $\{\lambda_{k,1}, \lambda_{k,2}\}$ or the penalty factor $\rho$ in the outer loop. Specially, the penalty factor is decreased with a constant $c$ ($0<c<1$) and  the dual variables are updated in the $i$-th iteration based on the following expressions
\begin{align}
\lambda_{k,1}^{(i+1)} \! &= \!\lambda_{k,1}^{(i)} 
\!+ \!\dfrac{1}{\rho^{(i)}}\!\left(x_k^{(i)} \!-\! \left(\bm{w}_k^{(i)}\right)^H\!\!\bm{a}(\theta_k)\bm{a}^T(\theta_k)\bm{u}_k^{(i)}\!\right),\forall k,\\ 
\lambda_{k,2}^{(i+1)} \!& = \!\lambda_{k,2}^{(i)} 
\!+\!  \dfrac{1}{\rho^{(i)}}\!\left(y_k^{(i)} \!-\!\left(\bm{w}_k^{(i)}\right)^H\!\!\left(\bm{h}_{ts}+\bm{H}_{ts}\bm{\phi}_1^{(i)}\!\right)\right),\forall k.
\end{align}
We define the constraint violation indicator as 
\begin{align}
	{h}(\mathbb{X}) =& \max\left\{\left|x_k- \bm{w}_k^H\bm{a}(\theta_k)\bm{a}^T(\theta_k)\bm{u}_k\right|, \right.\nonumber\\
	&~~~~~~~\left. \left|y_k-\bm{w}_k^H\left(\bm{h}_{ts}+\bm{H}_{ts}\bm{\phi}_1\right)\right|, \forall k \right\}. \label{in:h}
\end{align} 
By comparing the value ${h}(\mathbb{X})$ with the tolerance of accuracy, we can determine the termination of the outer loop. According to the analysis in \cite{PDD1}, the proposed PDD-based algorithm  converges to a stationary point of the communication SINR maximization problem (\ref{pb:o}).

Regarding the computational complexity, the proposed PDD-based algorithm has double loops, where the maximum iteration numbers of the outer loop and inner loop are $I_{\max}^o$ and $I_{\max}$, respectively. In  each iteration of the inner loop, we need to perform four steps. The computational complexities of four steps are  $\mathcal{O}\left( K\left(\! M^2( N_1 +1) +\log \dfrac{I_0}{\epsilon} \right)\right)$, $\mathcal{O}\left(K\left( \! M^2 ( N_2 +1) + \log \dfrac{I_0}{\epsilon} \right)\right)$, $\mathcal{O}\left(N_1^2M+N_1N_2\right)$, and $\mathcal{O}\left(N_2^2M+N_1N_2\right)$, respectively, where $I_0$ is the initialized interval length and $\epsilon$ is the tolerance of accuracy for the  bisection search. Therefore, the total computational complexity for the proposed PDD-based algorithm is 
\begin{align}
	&\mathcal{O}\left(I_{\max}^o I_{\max}\left(K(N_1+N_2)M^2+2K\log\dfrac{I_0}{\epsilon} \right.\right.\nonumber\\
	&~~~~~~~~~~~~~~~~~~~~\left.\left. + N_1^2 M +N_2^2M+2N_1N_2\right)\right). \label{comp:pdd}
\end{align}

\section{Special Case Analysis and Low-Complexity Algorithm}
To gain more insights for the double-RIS-assisted communication radar coexistence system, we first analyze the relationship between the radar power budget and the interference from the radar to the communication receiver. Then, we consider two special cases: one with large radar power and the other with low radar power. Moreover, a low-complexity design algorithm is composed based on the solutions to the special cases.

\subsection{Effect of the Radar Power Budget}

To analyze the effect of the radar power budget $P_{\max}$, we first obtain the optimal radar beamforming design under the given RIS passive beamforming vectors.  To maximize the communication SINR in problem  (\ref{pb:o}), the radar transmit and receive beamforming vectors, i.e., $\bm{u}_k$ and $\bm{w}_k$, tend to be designed to avoid the interference from the radar to the communication receiver. Therefore, we have the following Lemma.
\begin{lem}\label{thm:opt}
	The optimal solution to problem (\ref{pb:o}) under the given passive RIS beamforming vectors, denoted by $\left\{\bm{w}_k^{R,\star}, \bm{u}_k^{R,\star} \right\}$, can be expressed as  
	\begin{align}
		\bm{u}_k^{R,\star} \!&= \!\left\{\!\!\!
		\begin{array}{l}
		 \sqrt{\hat{\gamma}_k}\bm{a}^*(\theta_k)  \!-\!\dfrac{\hat{\bm{h}}_{sr}^T\bm{e}^*_{k}   \hat{\bm{h}}_{sr}^H \bm{a}(\theta_k)^*}{ |  \hat{\bm{h}}_{sr}^H \bm{e}_{k}|^2 + \hat{\lambda}^{\star}}	\sqrt{\hat{\gamma}_k}\bm{e}_{k},    \\~~~~~~~~~~~~~~~~~~~~~~~~~~~~~~~\text{if }|  \hat{\bm{h}}_{sr}^H \bm{e}_{k}| \neq 0,\\
		  \sqrt{\hat{\gamma}_k}\bm{a}^*(\theta_k),  ~~~~~~~~~~~~~~~~\text{otherwise},
		\end{array} \right. \label{res:u}\\
		\bm{w}_k^{R,\star} &= \left(\sigma^2 \bm{I} + p^c \hat{\bm{h}}_{ts}\hat{\bm{h}}_{ts}^H\right)^{-1}\bm{a}(\theta_k)\bm{a}^T(\theta_k)\bm{u}_k^{R,\star}, \label{res:w}
	\end{align}
	where $\hat{\gamma}^r_k \triangleq  \dfrac{\gamma^r \sigma^{2}}{|\alpha_k|^2 }\left(1 -\dfrac{ p^c\bm{a}(\theta_k)^H \hat{\bm{h}}_{ts}\hat{\bm{h}}_{ts} ^H\bm{a}(\theta_k)}{\sigma^{2}+p^c \hat{\bm{h}}_{ts} ^H \hat{\bm{h}}_{ts} }\right)^{-1}$, $\bm{e}_{k} \triangleq \dfrac{\hat{\bm{h}}_{sr}-(\bm{a}(\theta_k)^T\hat{\bm{h}}_{sr})\bm{a}^*(\theta_k)}{||\hat{\bm{h}}_{sr}-(\bm{a}(\theta_k)^T\hat{\bm{h}}_{sr})\bm{a}^*(\theta_k)||}$, $\hat{\bm{h}}_{sr} \triangleq {\bm{h}}_{sr}+\bm{H}_{sr}^H\bm{\phi}_2^*$, $\hat{\bm{h}}_{ts} \triangleq {\bm{h}}_{ts}+\bm{H}_{ts}\bm{\phi}_1$, and $\hat{\lambda}^{\star}$ is the optimal Lagrange multiplier for constraint (\ref{con:o_radar_power}). $\hat{\lambda}^{\star}$ satisﬁes $\hat{\lambda}^{\star}\!\left(\sum\limits_{k=1}^K\left|\left|\bm{u}_k^{R,\star}\right|\right|^2 \!-\!P_{\max} \right)=0$.
	\begin{IEEEproof}
		Please refer to Appendix \ref{proof:opt}. 
	\end{IEEEproof}
\end{lem}
Then, the corresponding maximized communication SINR can be expressed as 
\begin{equation}
 	\text{SINR}^{c,R,\star}  \!=\! \dfrac{KLp^c\left|h_{tr}\!+\!\bm{g}_{tr}^H\bm{\phi}_1\!+\!\bm{f}_{tr}^H\bm{\phi}_2\!+\!\bm{\phi}_1^T\bm{H}_{tr}\bm{\phi}_2\right|^2}{KL\sigma^2+\sum_{k=1}^K\left|\hat{\bm{h}}_{sr}^H\bm{u}_k^{R,\star}\right|^2}.\!\!\!
\end{equation}
Assuming $|  \hat{\bm{h}}_{sr}^H \bm{e}_{k}| \neq 0, \forall k$ \footnote{We should note that this condition can be satisfied in most cases, therefore, in the following we discuss the results based on this condition.}, we have the following Theorem to show the relationship between the radar power budget and the communication SINR.
\begin{thm} \label{thm:power}
	The interference from the radar to the communication receiver decreases with  the radar power budget, and thus the communication SINR increases with  the radar power budget. Specially, when the radar power budget is sufficiently large, the interference from the radar to the communication receiver is zero and the communication SINR is expressed as 
	\begin{equation}
	\text{SINR}^{c,R,\star} = \dfrac{p^c \left|h_{tr}+\bm{g}_{tr}^H\bm{\phi}_1+\bm{f}_{tr}^H\bm{\phi}_2+\bm{\phi}_1^T\bm{H}_{tr}\bm{\phi}_2\right|^2}{\sigma^2}.\!\label{SINR_non}
	\end{equation}
	\begin{IEEEproof}
		Please refer to Appendix \ref{proof:power}. 
	\end{IEEEproof}
\end{thm}

Lemma \ref{thm:opt} gives the optimal radar beamforming vector design under the given RIS passive beamforming vectors. The vectors, i.e., $\bm{a}^*(\theta_k) $ and $\bm{e}_{k}$, determine the radar beamforming for direction $\theta_k$ together. Specifically,  the power allocated to the steering vector $\bm{a}(\theta_k)$ affects the radar SINR, and $\bm{e}_k$ is related to the interference  channel vector $\bm{h}_{sr}$ and is orthogonal to $\bm{a}^*(\theta_k)$. The power allocated to $\bm{e}_k$ is utilized for reducing the interference to the communication receiver. Therefore,  when the radar transmit power is sufficiently large,  the power can be allocated to $\bm{a}^*(\theta_k)$ to satisfy the radar SINR requirement, and to $\bm{e}_{k}$ for avoiding the interference to the communication receiver. In this case, the communication SINR is not influenced by the radar system.  

Motivated by Theorem \ref{thm:power},  we can consider two special cases: one with large radar power and the other with low radar power. In the former one, the interference from the radar  to  the communication receiver is zero, and in the latter one, the interference dominates the communication SINR performance.

\subsection{Special Case 1: Large Radar Power}
We first assume that the radar power budget $P_{\max}$ is sufficiently large, which is practical. Then, according to Theorem \ref{thm:power}, the interference from the radar to the receiver is zero. Therefore, the RIS passive beamforming vectors are designed for enhancing the communication signal only, i.e., to maximize the $\text{SINR}^{c,R,\star}$. Furthermore, based on Lemma \ref{thm:opt}, by ignoring the influence provided by the ``transmitter--RIS 1--RIS 2--receiver'' link\footnote{This assumption is reasonable according to the product-distance path loss model \cite{RIS-channel}, the channel gain is relatively low after twice reflections under most circumstances.}, we can obtain the optimal beamforming vectors in closed-form as presented in the following Lemma. 
\begin{lem} \label{lem:phi}
If we assume $||\bm{H}_{tr}|| \ll ||\bm{f}_{tr}||$, the optimal phase shifts of RISs can be given by 
	\begin{align}
		\phi_{1,n}^{R,\star} &= \exp\left(j\angle  {h}_{tr} - j\angle \bm{g}_{tr}(n)\right),~~\forall n,\label{phi_1_opt}\\
		\phi_{2,n}^{R,\star} &= \exp\left(j\angle {h}_{tr}-j\angle \bm{f}_{tr}(n)\right),~~\forall n.\label{phi_2_opt}
	\end{align}
	The corresponding  maximized communication SINR is given by
	\begin{equation}
		\!\!\!\text{SINR}^{c,R,\star}  \!\!=\!  \dfrac{p^c\left||h_{tr}|\!+\!\sum_{n=1}^{N_1}|\bm{g}_{tr}(n)|\!+\!\sum_{n=1}^{N_2}|\bm{f}_{tr}(n)|\right|^2 }{\sigma^2}. \!\label{SINR_ris}
	\end{equation}
	\begin{IEEEproof}
		It is readily to  prove this  Lemma  and we omit the details for brevity.
	\end{IEEEproof}
\end{lem}

Based on Lemma \ref{thm:opt} and Lemma \ref{lem:phi}, we can obtain the optimal solution  to the communication SINR maximization problem (\ref{pb:o}) without constraint (\ref{con:o_radar_power})  in closed-form.  To summarize, the optimal solutions of the radar beamforming design are given in (\ref{res:u}) and (\ref{res:w}) and  the optimal passive beamforming vectors at the RISs in this case are given  in (\ref{phi_1_opt}) and (\ref{phi_2_opt}).

Then, the sufficient condition  of this special case can be given in the following Theorem. 
\begin{thm}
    When the following condition is met, the radar power is large and this special case occurs.
    \begin{equation}
    \sum_{k=1}^K\left|\left|\bm{u}_k^{R,\star}\right|\right|^2  = \sum_{k=1}^K\hat{\gamma_k}\left( 1 -\dfrac{   |\hat{\bm{h}}_{sr}^H \bm{a}^*(\theta_k)|^2}{ |  \hat{\bm{h}}_{sr}^H \bm{e}_{k}|^2 }\right) \le P_{\max}. \label{con:energy}
    \end{equation}
    \begin{IEEEproof}
        This Theorem can be easily proved and we omit the details for brevity.
    \end{IEEEproof}
\end{thm}

Next, we further analyze this special case by comparing the performance of the proposed double-RIS-assisted communication radar coexistence system with that of the conventional communication radar coexistence system without RIS. The optimal radar beamforming design in Lemma \ref{thm:opt} and results in Theorem \ref{thm:power}  can also be applied to the conventional system without RIS  and the  corresponding communication SINR be expressed as
\begin{equation}
	\text{SINR}^{c,N,\star} = \dfrac{p^c\left|h_{tr}\right|^2}{\sigma^2}, \label{eq:sinr_c}
\end{equation}
where the interference from radar to the receiver also is zero. Then, the performance gap between the proposed system and the conventional system  is presented in the following Theorem.
\begin{thm} \label{thm:gap}
	In this special case, the performance of the double-RIS-assisted communication radar coexistence system is higher than that of the conventional communication radar coexistence system  without RIS, and the performance gap is given by
\begin{align}
&\Delta \text{SINR}^{c} ~~~~~~~~~~~~~~~~~~~~~~~~~~~~~~~~~~~~~~~~~~~~~~~~~~~~~~\nonumber\\
&~= \!\dfrac{p^c\left||h_{tr}|\!+\!\sum\limits_{n=1}^{N_1}|\bm{g}_{tr}(n)|\!+\!\sum\limits_{n=1}^{N_2}|\bm{f}_{tr}(n)|\right|^2\!\!-\!\left|h_{tr}\right|^2}{\sigma^2} \!>\! 0.\label{res:gap}
\end{align}
	Assume that the RIS-related links are statistically independent and follow the Rayleigh distribution, i.e., $\bm{h}_{t1}\sim \mathcal{CN}\left(0, \varrho_{t1} \bm{I} \right)$, $\bm{h}_{1r}\sim \mathcal{CN}\left(0, \varrho_{1r} \bm{I} \right)$, $\bm{h}_{t2}\sim \mathcal{CN}\left(0, \varrho_{t2} \bm{I} \right)$, and $\bm{h}_{2r}\sim \mathcal{CN}\left(0, \varrho_{2r} \bm{I} \right)$.  As  $\min(N_1,N_2) \rightarrow \infty$, we have
	\begin{equation}
		\Delta \text{SINR}^{c} \rightarrow \dfrac{p^c}{\sigma^2} \left(N_1 \dfrac{\pi \varrho_{t1}\varrho_{1r}}{4}+N_2 \dfrac{\pi \varrho_{t2}\varrho_{2r}}{4}\right)^2.\label{res:gap_N}
	\end{equation}
	\begin{IEEEproof}
		The performance gap (\ref{res:gap}) is readily to obtain based on (\ref{SINR_ris}) and (\ref{eq:sinr_c}). Besides, we have $\sum_{n=1}^{N_1}|\bm{g}_{tr}(n)| = \sum_{n=1}^{N_1}|\bm{h}_{t1}(n)||\bm{h}_{1r}(n)|$. As  $N_1  \rightarrow \infty$, we have $\dfrac{\sum_{n=1}^{N_1}|\bm{h}_{t1}(n)||\bm{h}_{1r}(n)|}{N_1} \rightarrow  \dfrac{\pi \varrho_{t1}\varrho_{1r}}{4} $. Therefore, equation (\ref{res:gap_N}) can be achieved. The proof is thus completed.
	\end{IEEEproof}
\end{thm}

Theorem \ref{thm:gap} verifies that our proposed double-RIS-assisted communication radar coexistence  system outperforms the conventional system with assuming sufficiently large radar transmit power. Moreover, the performance gap  quadratically increases with the  number of reflecting elements under the assumption of Rayleigh fading channels. 

\subsection{Special Case 2: Low Radar Power}
In this special case, the radar has low transmit power budget $P_{\max}$. According to Theorem \ref{thm:power}, the interference is high and dominates the  communication SINR performance. Thus, we focus on reducing the interference between the communication transmitter and the radar as well as that between the radar and the communication receiver by optimizing $\bm{\phi}_1$ and $\bm{\phi}_2$, as 
\begin{subequations}\label{pb:phi}
	\begin{eqnarray}
		& \min\limits_{\bm{\phi}_1, \bm{\phi}_2} &  {\left|\left|\bm{h}_{sr}^H+\bm{\phi}_2^T\bm{H}_{sr}\right|\right|^2}+ \left| \left|\bm{h}_{ts}+\bm{H}_{ts}\bm{\phi}_1 \right|\right|^2,\\
		&\text{s.t.}& |\bm{\phi}_1(n)| = 1, |\bm{\phi}_2(n)| = 1,~\forall n.
	\end{eqnarray}
\end{subequations}
This problem aims at minimizing the power of the interferences and it can be solved directly by using the BCD method, where we optimize $\bm{\phi}_1(n),\forall n$ and $\bm{\phi}_2(n),\forall n$ sequentially. In each iteration, the solutions for $\bm{\phi}_1(n)$ and $\bm{\phi}_2(n)$ can be obtained similar to (\ref{res:phi_1}) and (\ref{res:phi_2}), and thus the details are omitted for brevity. After obtaining the beamforming vectors at the RISs, the radar transmit and receive beamforming vectors can be designed according to Lemma \ref{thm:opt}.

\subsection{Low-Complexity Algorithm}

Combining these two cases, we can design a low-complexity algorithm for solving problem (\ref{pb:o}) as presented in Algorithm \ref{ALG3}. Firstly, we obtain the communication SINR in two cases, denoted by $\text{SINR}^{c,H}$ and $\text{SINR}^{c,L}$. In the case of large radar power, the joint beamforming design can be obtained by Lemma \ref{thm:opt} and Lemma \ref{lem:phi}. In the case of low radar power, the BCD method can be adopted to obtain the solution of the beamforming vectors. Then, by comparing values of $\text{SINR}^{c,H}$ and $\text{SINR}^{c,L}$, we determine the ﬁnal communication SINR.  The corresponding  computational complexity of Step 2--Step 4 is $\mathcal{O}\left(M(N_1+N_2+K)+K\log\dfrac{I_0}{\epsilon}\right)$ and that of Step 6--Step 8 is $\mathcal{O}\left(I_{\max}^L(N_1^2+N_2^2)+M(N_1+N_2+K)+K\log\dfrac{I_0}{\epsilon}\right)$ with $I_{\max}^L$ being the maximum number of iteration for solving problem (\ref{pb:phi}). Based on the complexity mentioned above, Algorithm 3 can provide lower complexity than the PDD-based algorithm, and their performance will be verified via simulations.

{\begin{algorithm}[t]
		\caption{Proposed Low-Complexity Algorithm.}
		\label{ALG3}
		\DontPrintSemicolon
		\% Large radar power case\;
		Obtain $\bm{\phi}_1$ and $\bm{\phi}_2$ according to (\ref{phi_1_opt}) and (\ref{phi_2_opt});\;
		Obtain $\bm{u}_k$  and $\bm{w}_k$ according to (\ref{res:u}) and (\ref{res:w});\;
		Calculate $\text{SINR}^{c,H}$;\;
		\% Low radar power case\;
		Obtain $\bm{\phi}_1$ and $\bm{\phi}_2$ by solving problem (\ref{pb:phi});\;
		Obtain $\bm{u}_k$  and $\bm{w}_k$ according to (\ref{res:u}) and (\ref{res:w});\;
		Calculate $\text{SINR}^{c,L}$;\;
		\eIf{$\text{SINR}^{c,H}\ge \text{SINR}^{c,L}$}{
		\textbf{Output}  $\text{SINR}^{c,H}$.\;}{
		\textbf{Output}  $\text{SINR}^{c,L}$.\;
		}
	\end{algorithm}
}

\section{Simulation Results}
In this section, we verify the effectiveness of the proposed  joint beamforming design algorithms in the double-RIS-assisted communication radar coexistence system.

\subsection{Simulation Setup}

\begin{figure}[b]
	\vspace{-4ex}
	\centering
	\includegraphics[width=0.4\textwidth]{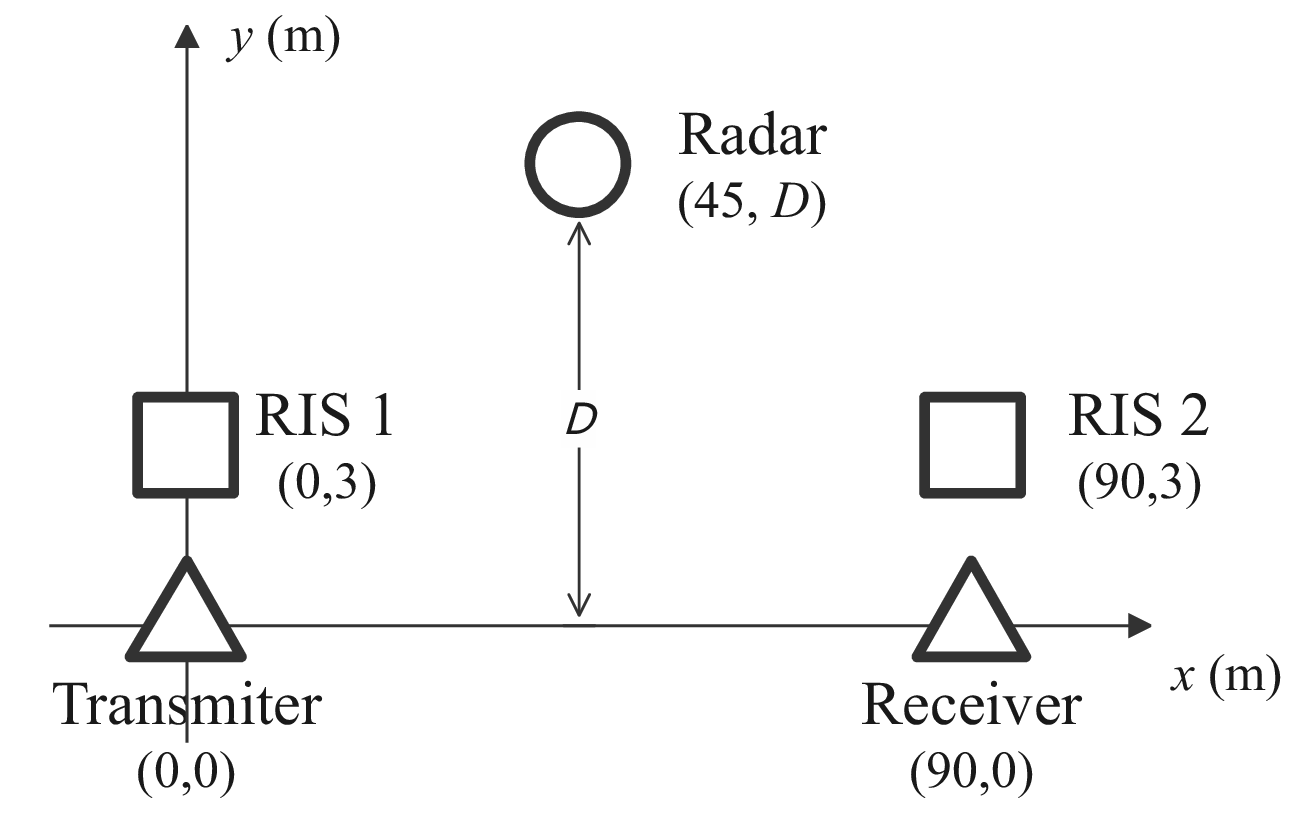}
	\vspace{-1ex}
	\caption{Simulation layout (top view).}
	\label{fig:sim}
\end{figure}
We consider a double-RIS-assisted communication radar coexistence system as shown in Fig. \ref{fig:sim}. The central points of the transmitter, receiver, RIS 1, RIS 2, and radar are located at (0, 0), (90, 0), (0, 3), (90, 3), and (45, $D$) in meter (m), respectively. The default  setting of $D$ is 20. Both RISs are equipped with 100 reflecting elements, i.e., $N_1=N_2 = 100$, and the numbers of transmit and receive antennas for the radar are 12, i.e., $M=12$. The antenna spacing is set as ${\lambda_o}/{2}$, i.e., $\Delta = {\lambda_o}/{2}$. The frequency considered in the simulation is 2.4 GHz. The large-scale fading of the wireless  channel is modelled as 
\begin{equation}
	L(d) = 32.6 + 36.7 \log(d), 
\end{equation}
where $d$ denotes the individual link distance in meter (m). As for the small-scale fading, it follows Rician fading, which is modelled as 
\begin{equation}
	\bm{H} =   \sqrt{\dfrac{\varepsilon}{1+\varepsilon}} \bm{a}_{M^r}(\theta^r)\bm{a}_{M^t}(\theta^t)^T +\sqrt{\dfrac{1}{1+\varepsilon}}\bm{H}_0, \label{channel_model}
\end{equation}
where $\varepsilon$ is the Rician factor, $M^r$ is the number of receive antennas (or reflecting elements), $M^t$ is the number of transmit antennas (or reflecting elements), $\theta^r$ and $\theta^t$ are corresponding azimuth angles, and $\bm{H}_0$ is the non-line-of-sight component  whose entries follow the distribution $\mathcal{CN}(0, 1)$.  We set $\varepsilon=9$ dB for the communication-related links, i.e., $h_{tr}$, $\bm{h}_{t1}$, $\bm{H}_{12}$, $\bm{h}_{1r}$, $\bm{h}_{t2}$, and $\bm{h}_{2r}$, and set $\varepsilon=3$ dB for the interference-related links, i.e., $\bm{h}_{ts}$,  $\bm{h}_{sr}$,  $\bm{H}_{1s}$, and  $\bm{H}_{s2}$. Besides, equation (\ref{channel_model}) becomes a Rayleigh fading channel model when $\varepsilon$ is set as 0. The transmit power of the  communication transmitter is set as $p^c = 0.1$ W  and the noise power is $10^{-13}$ W.  As for the radar, there are $8$ detection directions in the radar system, that is $-\dfrac{\pi}{3},-\dfrac{\pi}{4},\cdots,\dfrac{\pi}{4}$, and the PRI is 10, i.e., $L=10$. The ratio of  $|\alpha_k|^2$ to noise power is set as $\dfrac{|\alpha_k|^2}{\sigma^2} = -12 $ dB, $\forall k$. The total transmit power for 8 directions is set as $10$ W and the SINR requirement is $10$ dB for each direction.

\subsection{Algorithm Investigation}

\begin{figure}[t]
		\centering
		\includegraphics[width=0.4\textwidth]{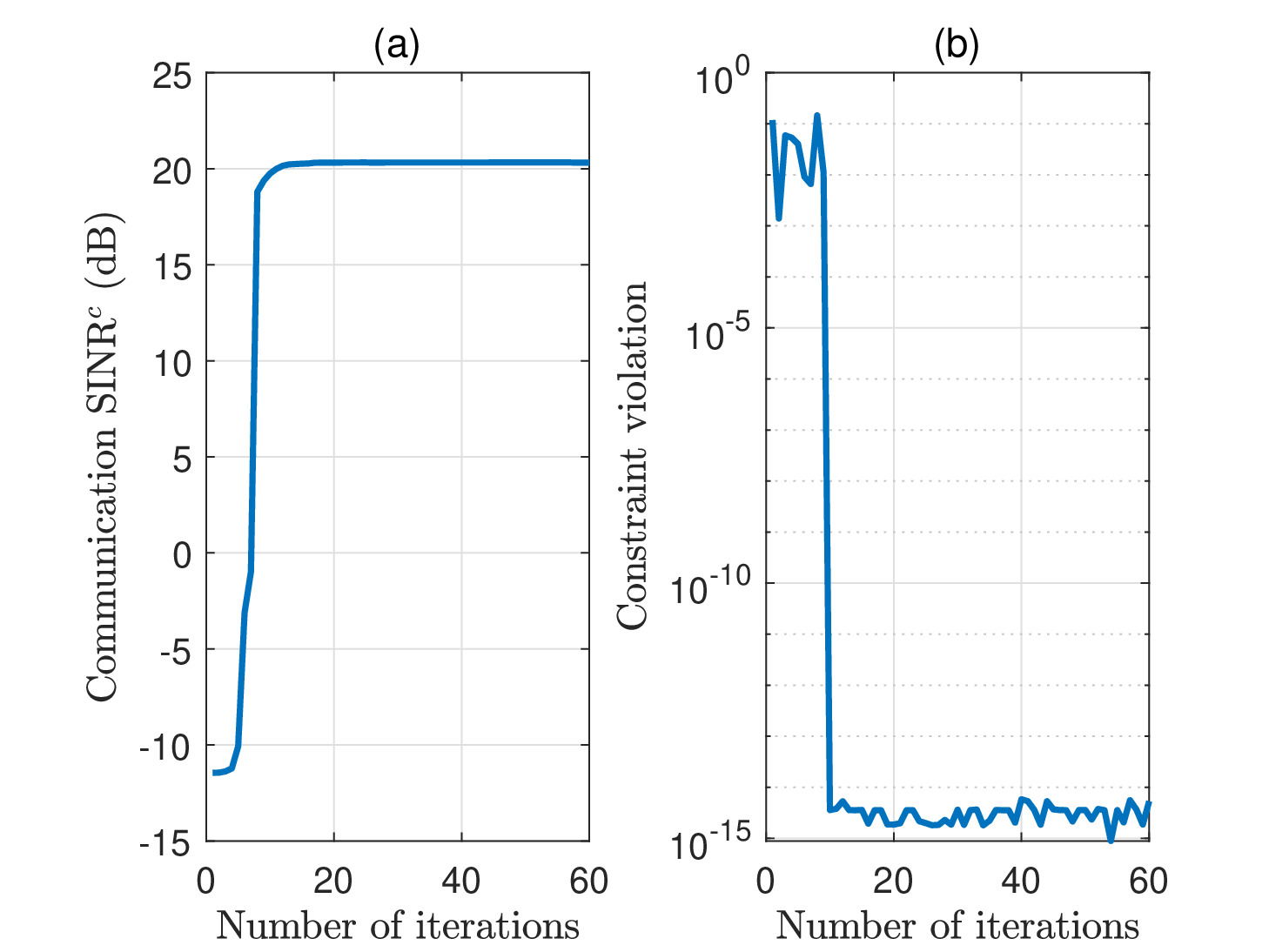}
		\vspace{-2ex}
		\caption{Convergence behavior of the proposed PDD-based algorithm.}
		\vspace{-2ex}
		\label{fig:Convergence}
\end{figure}

\begin{figure}[t]
	\centering
	\includegraphics[width=0.4\textwidth]{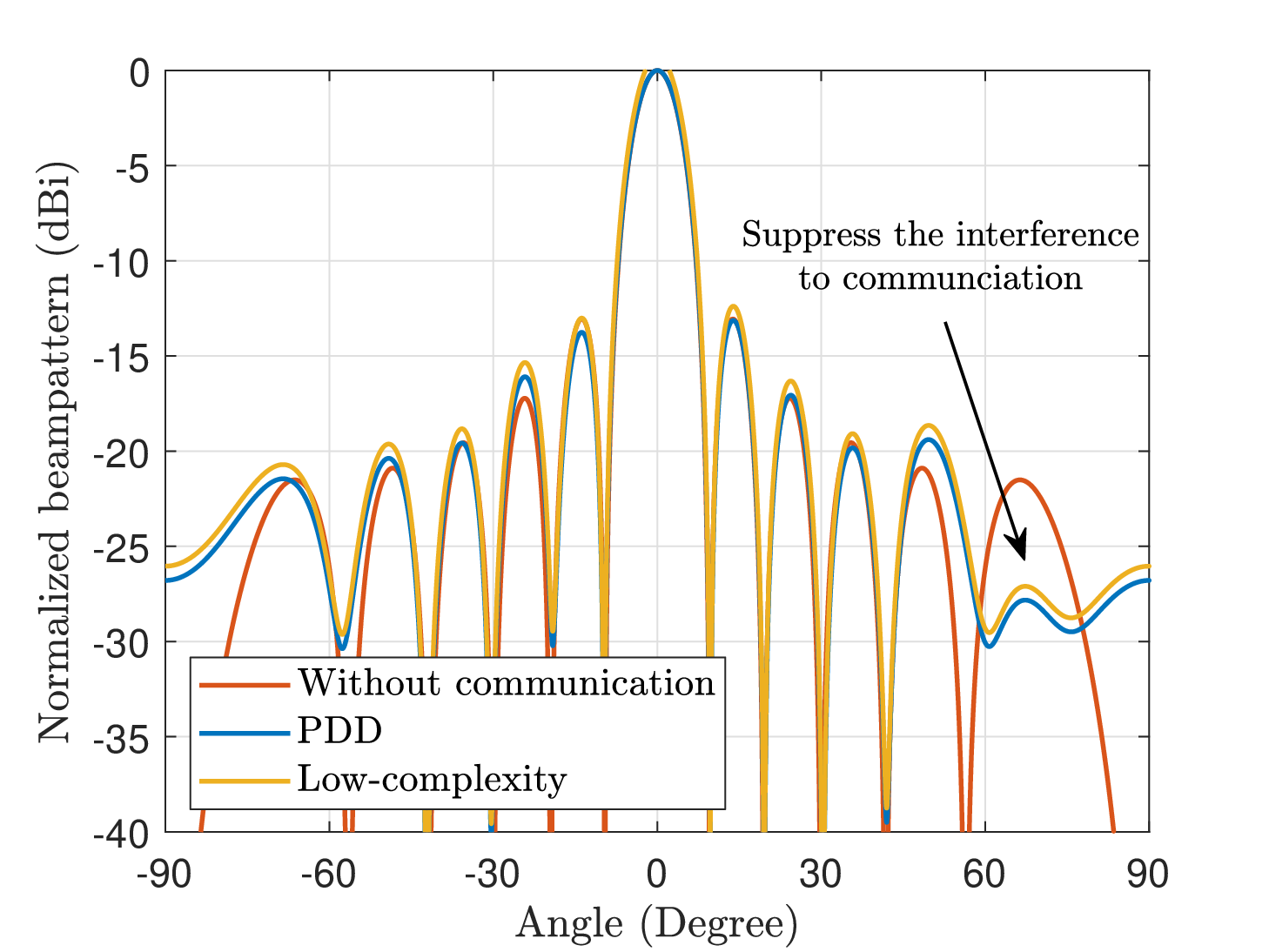}
	\vspace{-2ex}
	\caption{Normalized beampattern comparison.}
	\vspace{-2ex}
	\label{fig:beam}
\end{figure}

First of all, we show the convergence behavior of the proposed PDD-based algorithm versus the iteration number of the outer loop as shown in Fig. \ref{fig:Convergence}. From the curves, we observe that the proposed algorithm achieves the convergence within about 10 iterations, which demonstrates a fast convergence rate. Meanwhile, the value of the constraint violation indicator, referring to ${h}(\mathbb{X})$ in (\ref{in:h}), decreases to around $10^{-10}$. It confirms that the solution obtained by the PDD-based algorithm is feasible to the original problem. Furthermore, as the iteration continues, the  value of the constraint violation indicator is less than  $10^{-14}$.

Next, we plot the normalized beampatterns of the analyzed algorithms when the detection direction is $0$ degree and compare them with the beampattern  of a conventional radar system  without communication, i.e., $\bm{u}_k = \bm{a}^*(\theta_k)$. We observe that the main lobes of these approaches are almost the same, which indicates that our radar beamforming design has little loss. Besides, the side lobes obtained by the proposed two algorithms are different from that of  $\bm{a}^*(\theta_k)$ for better suppressing the interference from the radar to the communication receiver.

\subsection{Performance Comparison}
To show the performance improvement, we consider the following two benchmark schemes.
\begin{itemize}
	\item Random phase design: the phase shift of reflecting elements is randomly generated by following the uniform distribution within $[0, 2\pi)$ and the radar beamforming vectors are designed according to (\ref{res:u}) and (\ref{res:w}).
	\item Conventional system: there is no RIS in the  radar and communication coexistence system and the radar beamforming vectors are designed according to (\ref{res:u}) and (\ref{res:w}).
\end{itemize}

\begin{figure}[t]
    \centering
	\includegraphics[width=0.4\textwidth]{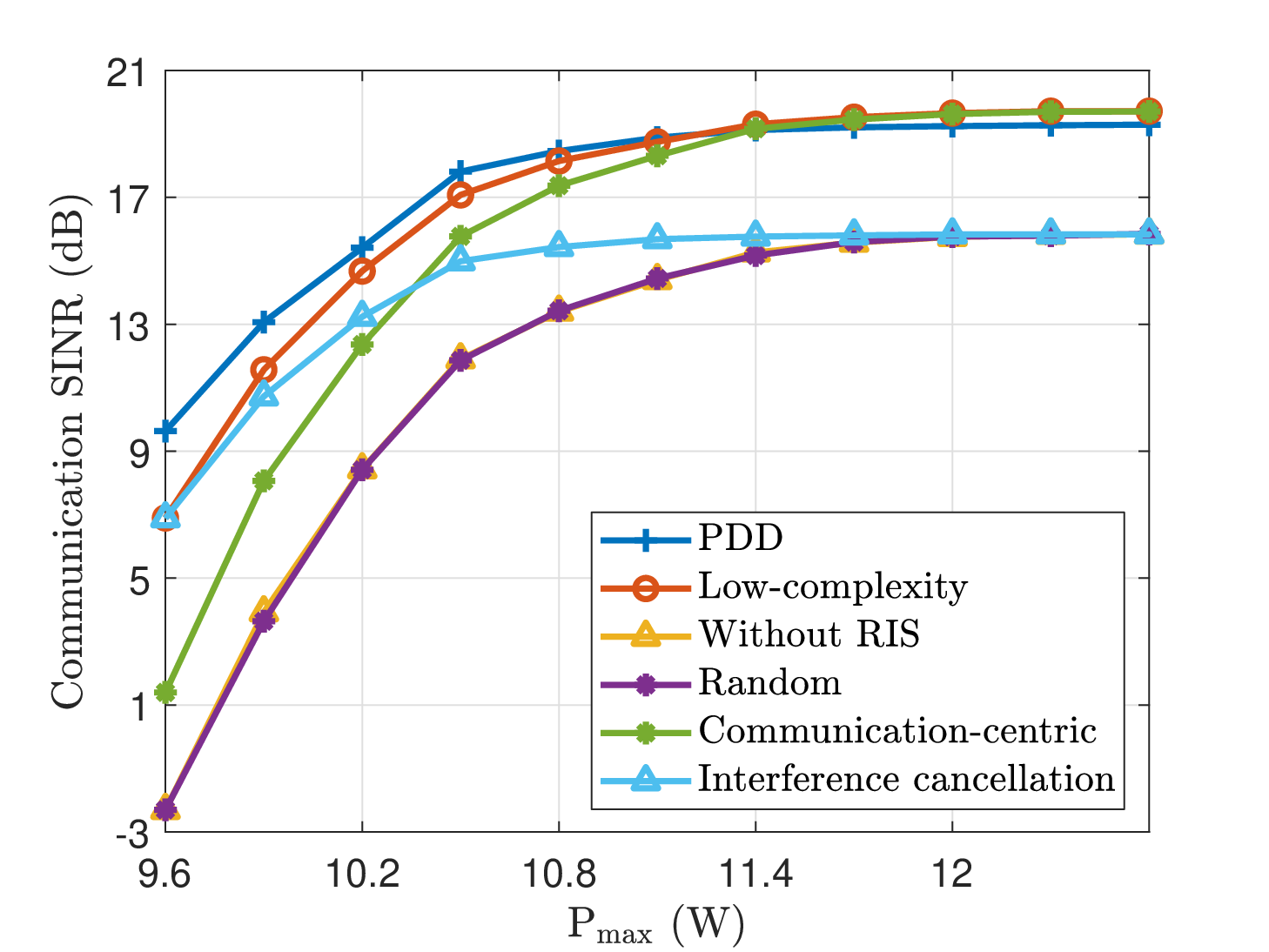}
	\vspace{-2ex}
	\caption{Communication SINR versus  $P_{\max}$.}
	\vspace{-2ex}
	\label{fig:pmax}
\end{figure}

\begin{figure}[t]
		\centering
		\includegraphics[width=0.4\textwidth]{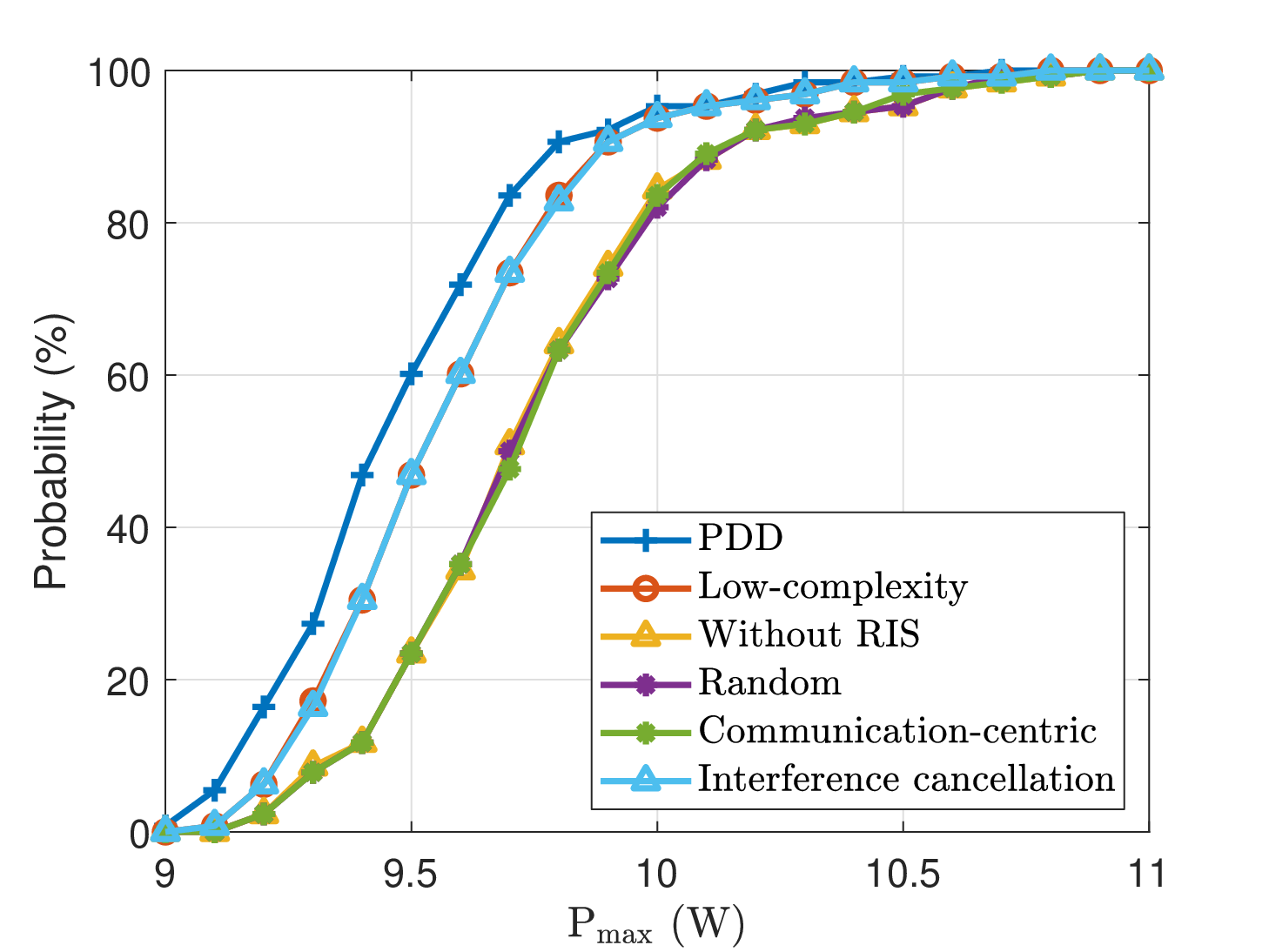}
		\vspace{-2ex}
		\caption{Successful probability of a feasible solution versus  $P_{\max}$.}
		\vspace{-2ex}
		\label{fig:feasible}
\end{figure}

\begin{table*}[t]
	\centering
	\caption{Performance comparison ($N_1 = 40$)} \label{tab:Pc}
	\vspace{-2ex}
	 \setlength{\tabcolsep}{3mm}{
	\begin{tabular}{|c|c|c|c|c|c|}
	\hline
	Algorithm & PDD-based & Low-complexity & \makecell[c]{Communication-\\centric}  & \makecell[c]{Interference\\cancellation}  & Random \\
	\hline\hline
	Running time (s)& 0.7135 & 0.0237& 0.0105& 0.0132& 0.0101\\
	\hline 
	Communication SINR (dB) & 11.37 & 10.34 & 8.54 & 8.97 & 6.66\\
	\hline
	\end{tabular}}
\end{table*}

Fig. \ref{fig:pmax} plots the effect of the total radar transmit power, i.e., $P_{\max}$, on the communication SINR for  the analyzed schemes. Besides, we also provide the results of the communication-centric design and the interference cancellation design. The former  aims at maximizing communication SINR without considering the radar, and the latter aims at minimizing the mutual interference. Note that the communication-centric design is given in the special case of  large radar power and the interference cancellation design is given in the special case of low radar power.  First of all, the communication SINRs of the four schemes all increase with $P_{\max}$. 
Secondly, the communication-centric design shows the high communication SINR under a large radar power budget, and the interference cancellation design shows the high communication SINR under a low radar power budget. Since the low-complexity algorithm is the combination of two designs, it achieves a high communication SINR  with any value of radar power budget. 
Thirdly, we can see that the PDD-based algorithm achieves the best performance under low $P_{\max}$ while the low-complexity algorithm achieves the best performance under high $P_{\max}$. We should note that the average minimum $P_{\max}$ that satisfies the condition for large radar power  in equation (\ref{con:energy}) is 11.07 W according to our simulation results. This value is close to the intersection point of the ``PDD'' and ``Low-complexity'' curves. When $P_{\max}$ is higher than 11.07 W, the solution obtained by the low-complexity algorithm is almost optimal according to Lemma \ref{thm:opt} and Lemma \ref{lem:phi}. This result also demonstrates the high performance of the proposed two algorithms. 
Finally, the performance of the random design is similar to that of the conventional system since deploying RISs with random phase shifts enhances the communication signal and interference simultaneously. 

In the meanwhile, as shown in Fig. \ref{fig:feasible}, we study that the effect of $P_{\max}$ on the  successful probability of a feasible solution. It can be seen that the PDD-based algorithm achieves the best successful probability followed by the proposed low-complexity algorithm, which validates the effectiveness of the proposed algorithms. Besides, the probability increases with  $P_{\max}$ since the radar SINR requirement is easier to be satisfied.

The comparison of the running time and the achieved communication SINR between the proposed and benchmark algorithms with $N_1=40$ is shown in Table \ref{tab:Pc}.  We can see that all algorithms have a low running time and the proposed algorithms significantly outperform the benchmark algorithms. Besides, the low-complexity algorithm costs less running time than the PDD-based algorithm.

\begin{figure}[t]
	\vspace{-3ex}
	\centering
	\includegraphics[width=0.4\textwidth]{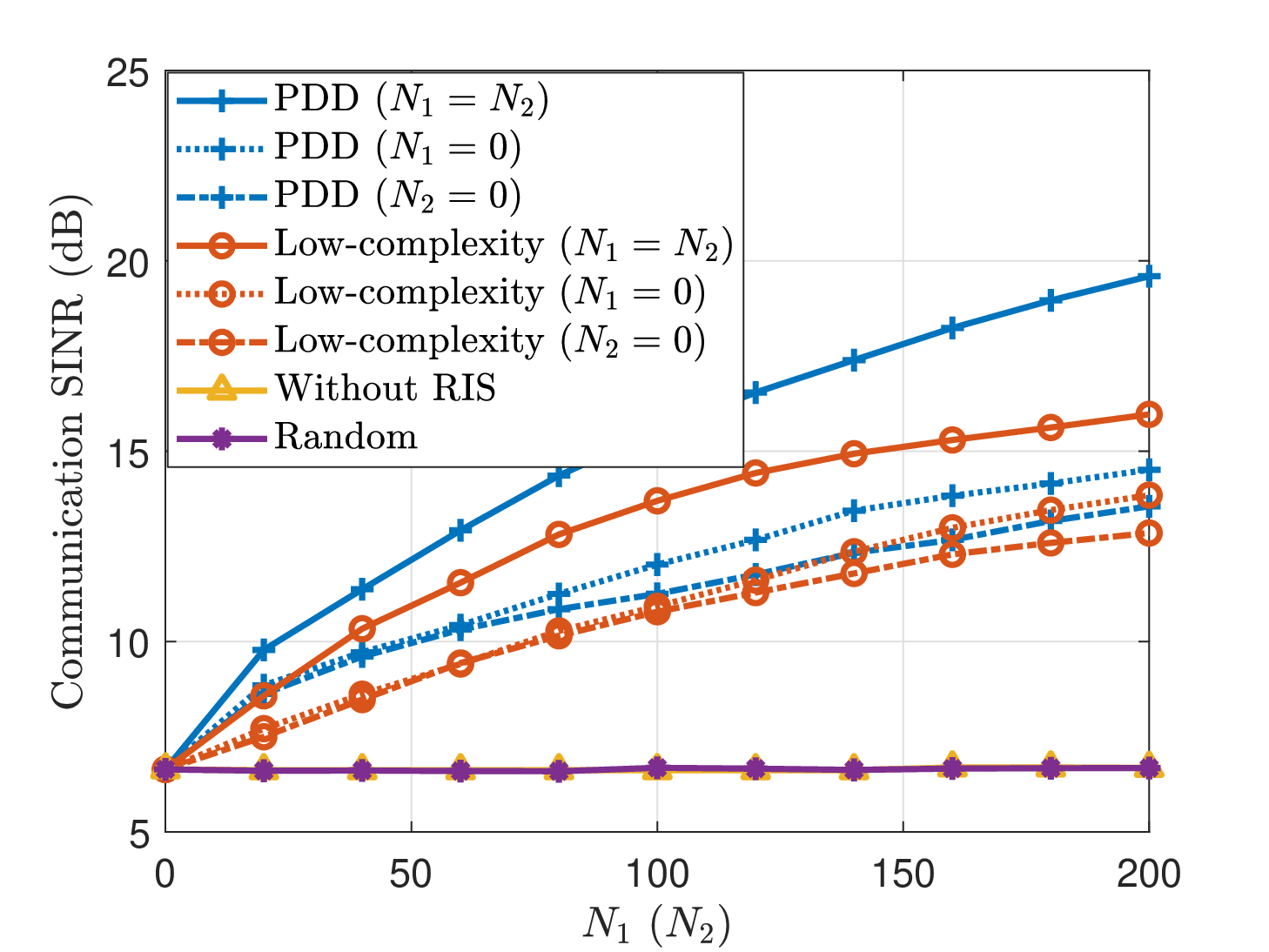}
	\vspace{-2ex}
	\caption{Communication SINR versus $N_1$ ($N_2$).}
	\vspace{-2ex}
	\label{fig:irsnum}
\end{figure}

In Fig. \ref{fig:irsnum}, we investigate the effect of the number of reflecting elements. There are three curves for the PDD-based algorithm or low-complexity algorithm, namely, $N_1=N_2$ (i.e., two RISs), $N_1=0$ (i.e., only one RIS deployed near the receiver), and $N_2=0$ (i.e., only one RIS deployed near the transmitter). We can see that  the double-RIS-assisted system with the PDD-based algorithm (or the low-complexity algorithm) outperforms the conventional systems with different values of $N_1$ (or $N_2$) and the performance gap between them increases with $N_1$ (or $N_2$). With more reflecting elements, the passive beamforming gain brought by the RIS for communication  increases and the interference can be further suppressed. This demonstrates that introducing RISs to the  communication radar coexistence system improves performance. Furthermore, compared with the single-RIS-assisted system, the interference can be better suppressed in the proposed double-RIS-assisted system, which shows the necessity of deploying two RISs.  Besides, it can be seen that  placing a  RIS near the  receiver is more efficient  than placing a RIS near the transmitter since the RIS can directly reduce the interference from the radar to the communication receiver.

\begin{figure}[t]
    \vspace{-3ex}
	\centering
	\includegraphics[width=0.4\textwidth]{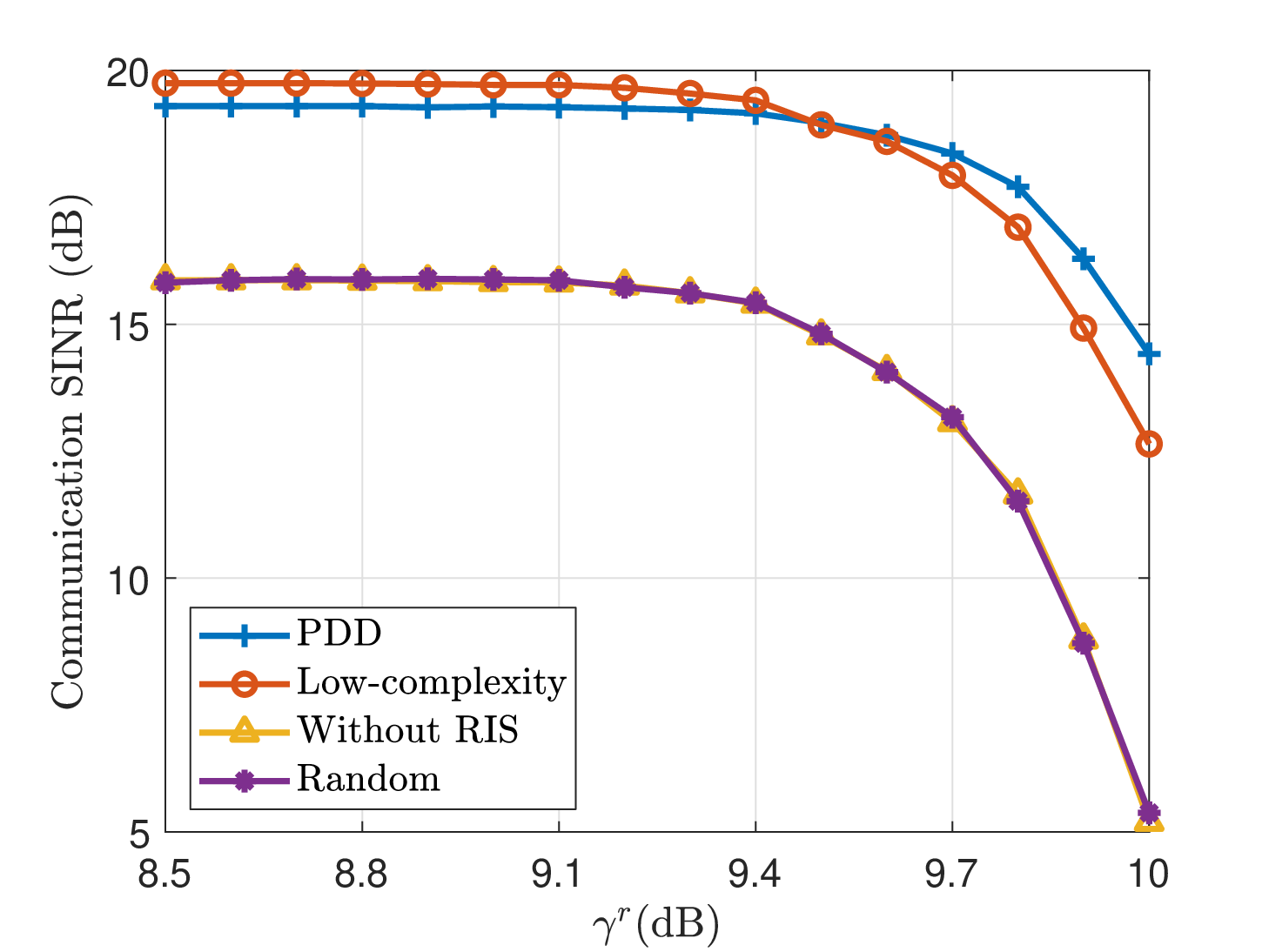}
	\vspace{-2ex}
	\caption{Communication SINR versus  $\gamma^r$.}
	\label{fig:gammar}
	\vspace{-2ex}
\end{figure}

In Fig.  \ref{fig:gammar}, we show the communication SINR versus the radar SINR requirement, i.e., $\gamma^{r}$. We can see that communication SINR remains almost unchanged when  $\gamma^r$ is low but decreases with  $\gamma^r$ when $\gamma^r$ is high. It is because that when $\gamma^r$ is low, the radar transmit power is adequate and the interference can be well suppressed. As  $\gamma^r$ increases, more  radar transmit power is utilized to maintain the radar SINR and less power is utilized to suppress the interference from the radar to the communication receiver, which leads to the decrease of the communication SINR. Moreover, the two proposed algorithms also show better performance than other algorithms.

\begin{figure}[t]
	\centering
	\includegraphics[width=0.4\textwidth]{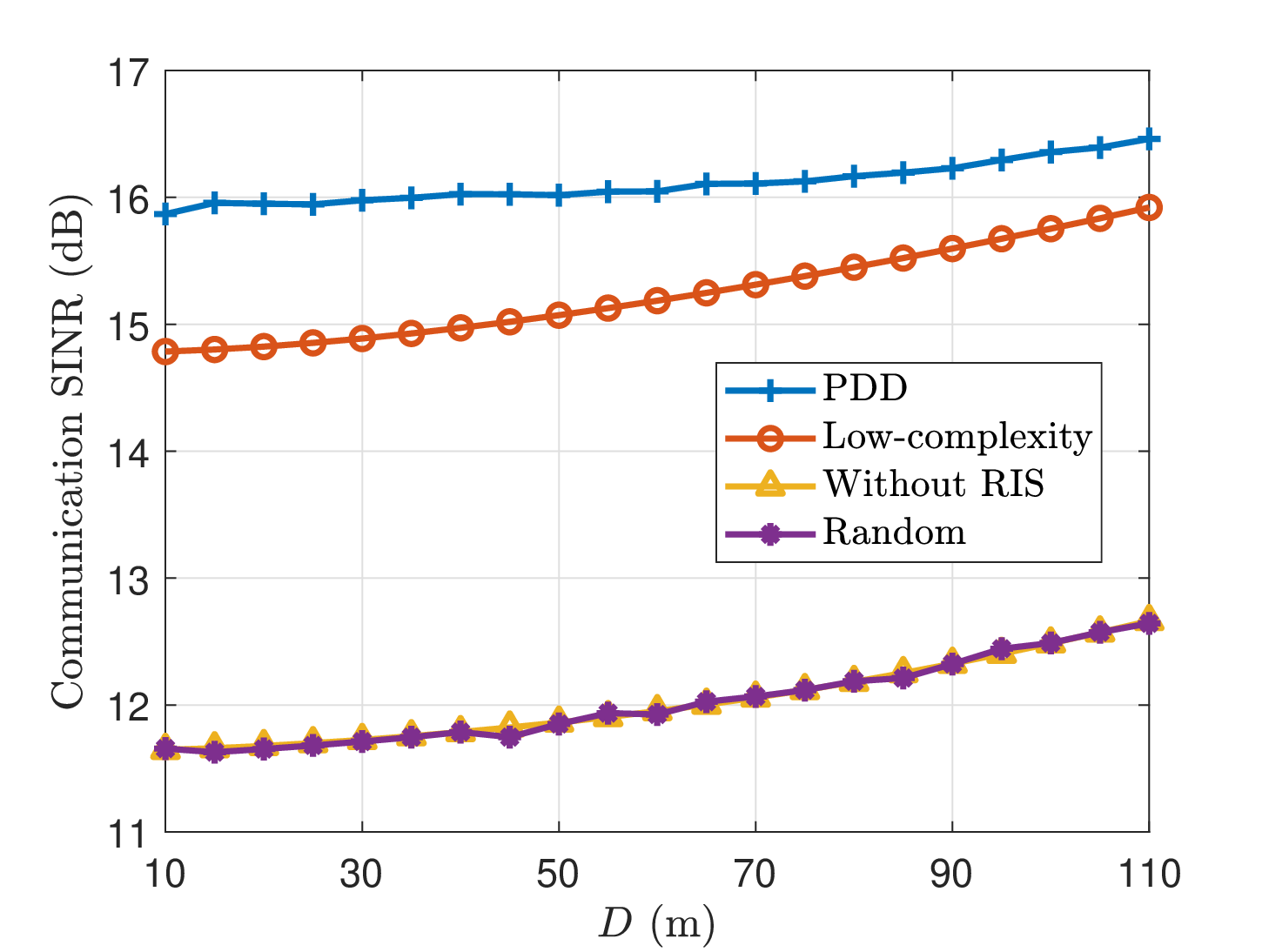}
	\vspace{-2ex}
	\caption{Communication SINR versus  $D$ under the assumption of Rayleigh fading channels.}
	\vspace{-2ex}
	\label{fig:distance2}
\end{figure}

\begin{figure}[t]
    \centering
		\includegraphics[width=0.4\textwidth]{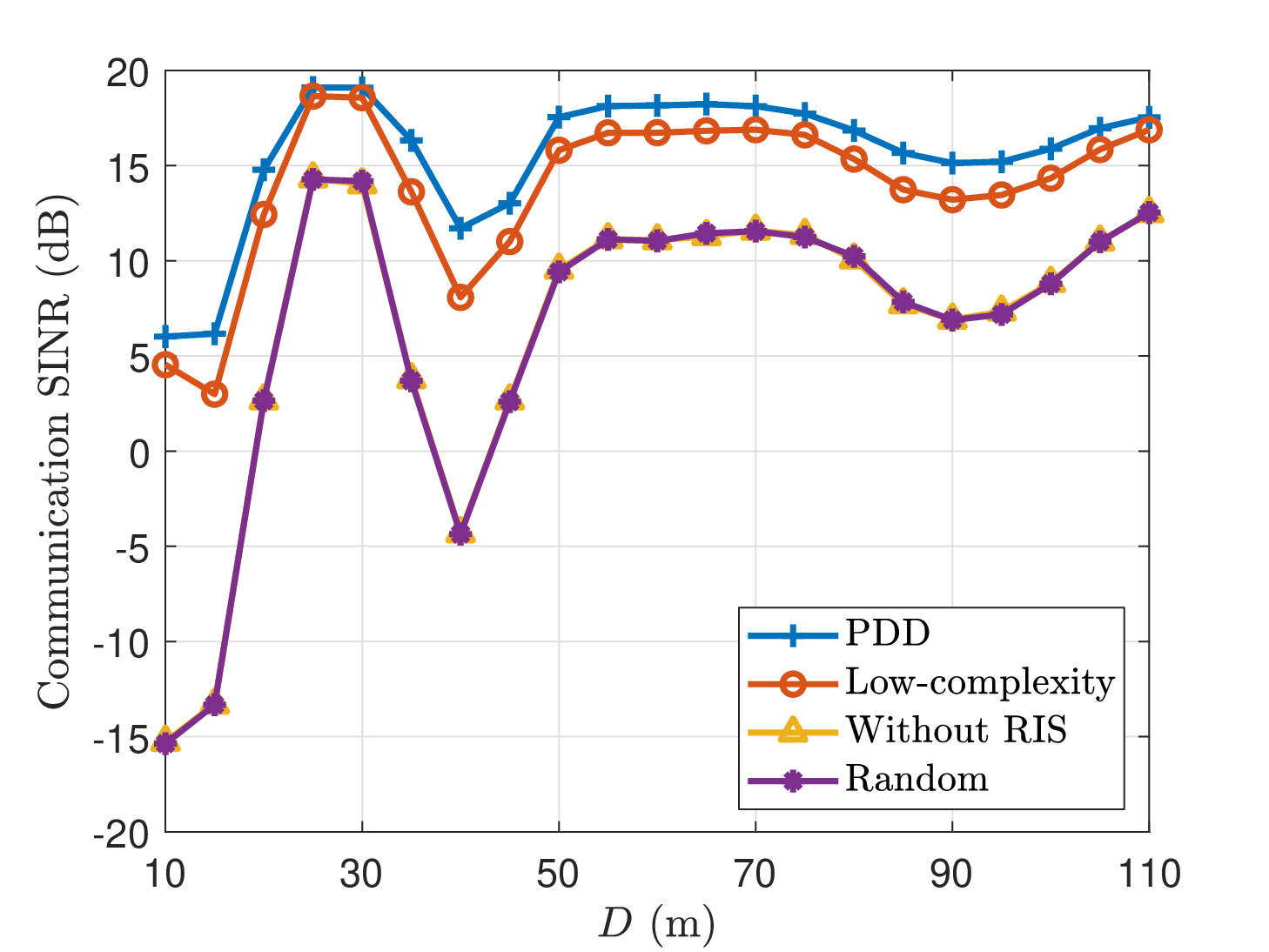}
		\vspace{-2ex}
		\caption{Communication SINR versus  $D$ under the assumption of Rician fading channels.}
		\vspace{-2ex}
		\label{fig:distance}
\end{figure}

The impact of the radar location, i.e. $(45,D)$, on the communication SINR under both Rayleigh fading and Rician fading channels are shown in Fig. \ref{fig:distance2} and Fig. \ref{fig:distance}, respectively. For Rayleigh fading channels, the interference channel between the radar and the communication transmitter/receiver is dominated by the distance, i.e., $\sqrt{45^2+D^2}$.  As $D$ increases, the large-scale fading increases and the interference channel gain decreases, which leads to the increase of communication SINR. Moreover, the performance difference between the PDD-based algorithm and low-complexity algorithm decreases with $D$. It is because that the effect of interference becomes smaller with $D$ and thus the radar transmit power becomes  sufficiently large. Regarding the Rician fading channels, the interference channel is affected by the distance and azimuth angle. The former causes that the communication SINR has an increasing trend with $D$ approximately as shown in Fig. \ref{fig:distance}, and the latter influences the small-scale fading and results in the fluctuations of the communication SINR. Specifically, when $D$ is in the range of $[25,30]$, the communication performance is higher than that of other values of $D$. It is because the radar beamforming vector is orthogonal to the interference channel in this case and thus the interference between the radar and the communication receiver can be well suppressed.

\begin{figure}[t]
	\centering
	\includegraphics[width=0.4\textwidth]{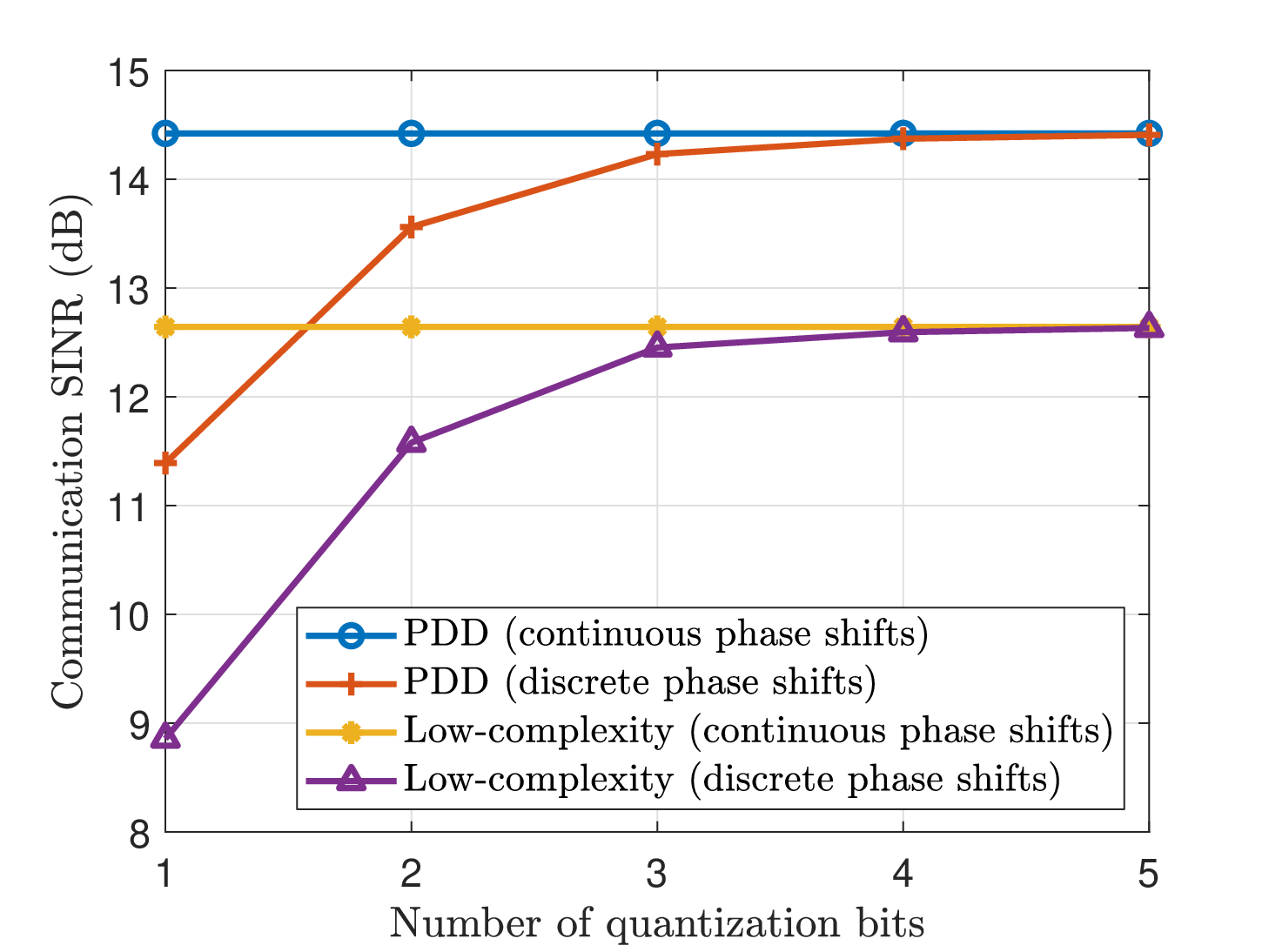}
	\vspace{-2ex}
	\caption{Communication SINR versus the number of quantization bits $b$.}
	\vspace{-2ex}
	\label{fig:bit}
\end{figure}

Considering the finite phase shift of the RIS in practice, we can discretize the continuous phase shifts of the RIS. Let $b$ denote the number of quantization bits and then the set of discrete phase shifts is $\mathcal{M} = \left\{\dfrac{m\pi}{2^{b-1}} | m = 0,1,2,\cdots, 2^b-1\right\}$. Let $\{\bm{\phi}_1^{\star}(n),\bm{\phi}_2^{\star}(n)\}$ denote the solution obtained by the PDD-based algorithm (or the low-complexity algorithm). Then, the discrete phase shifts are given by 
\begin{align}
	\widehat{\bm{\phi}}_1(n) &= \mathop{\arg\min}_{\{\phi\in \mathcal{M}\}} \left\{ \left|\phi - \bm{\phi}_1^{\star}(n)\right| \right\},\\
	\widehat{\bm{\phi}}_2(n) &=  \mathop{\arg\min}_{\{\phi\in \mathcal{M}\}} \left\{ \left|\phi - \bm{\phi}_2^{\star}(n)\right| \right\}.
\end{align} 
The effect of the number of quantization bits on the communication SINR is shown in Fig. \ref{fig:bit}.  We can see that as the number of quantization bits increases, the communication SINR obtained by our proposed algorithms with discrete phase shifts increases and approaches the communication SINR under the continuous condition. When the number of quantization bits is more than $3$, the performance with discrete phase shifts is almost the same as that of continuous phase shifts. This verifies the effectiveness of our proposed algorithm in practice.

\section{Conclusion}
In this paper, we have investigated the double-RIS-assisted communication radar coexistence system to further enhance the communication signal and suppress the interference, where two RISs are placed near the communication transmitter and receiver, respectively. To this end, a communication SINR maximization problem has been formulated subject to constraints of the radar SINR requirement and the radar transmit power limitation. To deal with the nonconvex objective function and nonlinear constraint, a double-loop PDD-based algorithm has been proposed. Specifically, the inner loop solves the AL problem via the CCCP method while the outer loop updates the AL multipliers or the penalty factor. To gain more insights, we studied two special cases of large radar power and low radar power.  In the first case, we derived the optimal joint beamforming design and showed the performance gain by comparing the proposed system with the conventional system without RIS. In the second case, the mutual interference was minimized via the BCD method. By combining two cases, a low-complexity algorithm was developed. Finally, numerical results were presented to verify the effectiveness of the proposed system and two algorithms.

For simplicity, we only considered one transmitter and one receiver, and they are both equipped with a single antenna. In the general scenario, there may be multiple receivers, and the transmitter and receivers are usually equipped with multiple antennas. Thus, the joint design of the transmitter/receiver beamforming, the RIS passive beamforming, and the radar beamforming is a critical problem to investigate in the future.

\appendices

\section{Details of CCCP Algorithm} \label{app:step}
In this appendix, we show the detailed derivation of solving subproblems (\ref{pb:step1}) and (\ref{pb:u}).
\subsection{Optimal Solution to Subproblem  (\ref{pb:step1})}
The Lagrange function of subproblem (\ref{pb:step1}) can be expressed as 
\begin{align}
	\mathcal{L} =& |x_k - \bm{w}_k^H \bm{q}_k+\rho\lambda_{k,1}|^2+ |y_k-\bm{w}_k^H \bm{f} +\rho\lambda_{k,2}|^2 \nonumber\\
	&+\mu_k \left( \gamma^r\left({\sigma^2\left|\bm{w}_k\right|^2+p^c\left|y_k \right|^2}\right) +  |\alpha_k|^2 \left|x_k^{(j)}\right|^2\right. \nonumber \\
	&~~~~~~~~~~~~-\left.2|\alpha_k|^2 \Re\left(\left(x_k^{(j)}\right)^* x_k \right) \right),
\end{align}
where $\mu_k\ge 0$ is the Lagrange multiplier associated with constraint (\ref{con:radar_sin3}).  By examining the Karush-Kuhn-Tucker (KKT) conditions, we obtain
\begin{align}
	\dfrac{\partial \mathcal{L}}{\partial \bm{w}_k} =& -2\left(x_k - \bm{w}_k^H \bm{q}_k+\rho\lambda_{k,1} \right)^H \bm{q}_k   +2 \mu_k\gamma^r \sigma^2 \bm{w}_k \nonumber\\
	& - 2\left(y_k-\bm{w}_k^H \bm{f} +\rho\lambda_{k,2}\right)^H \bm{f}= 0,\\
	\dfrac{\partial \mathcal{L}}{\partial x_k} =& 2\left(x_k - \bm{w}_k^H \bm{q}_k+\rho\lambda_{k,1} \right) - 2 \mu_k|\alpha_k|^2  x_k^{(j)} = 0,\\
	\dfrac{\partial \mathcal{L}}{\partial y_k} =& 2\left(y_k-\bm{w}_k^H \bm{f} +\rho\lambda_{k,2}\right) + 2\mu_k \gamma^r p^c y_k = 0, \label{eq:kkt-mu}
\end{align}
\begin{align}
\mu_k \left( \gamma^r\left({\sigma^2\left|\bm{w}_k\right|^2+p^c\left|y_k \right|^2}\right) +  |\alpha_k|^2 \left|x_k^{(j)}\right|^2~~~~~~~~\right.&\nonumber\\
\left.- 2|\alpha_k|^2 \Re\left(\left(x_k^{(j)}\right)^* x_k \right) \right) = 0.&
\end{align}
Then, we can discuss the solution based on the value of $\mu_k^{\star}$ as follows.

\textbf{Case 1}: If $\mu_k^{\star}=0$, the optimal solution satisfies $ x_k^{\star}= \left( \bm{w}_k^{\star}\right) ^H \bm{q}_k-\rho\lambda_{k,1}$, $y_k^{\star} = \left(\bm{w}_k^{\star}\right)^H \bm{f} -\rho\lambda_{k,2}$, and equation (\ref{eq:kkt-mu}). With simple mathematical calculation, we can derive the following solution
\begin{align}
	\bm{w}_k^{\star} & = \left(\gamma^r\sigma^2 \bm{I}+\gamma^r p^c \bm{f} \bm{f}^H\right)^{-1} \nonumber \\
	&~~~~~~~~~~~~~~\times\left(\gamma^rp^c \rho \lambda_{k,2}^*\bm{f}+|\alpha_k|^2 \left(x_k^{(j)}\right)^*\bm{q}_k\right).
\end{align}

\textbf{Case 2}: If $\mu_k^{\star} \neq 0$, the optimal solution can be given by 
\begin{align}
	\bm{w}_k^{\star} & = \left(\dfrac{\mu_k^\star \gamma^r p^c }{1+\mu_k^\star \gamma^r p^c} \bm{f}\bm{f}^H + \mu_k^{\star}\gamma^r \sigma^2 \bm{I}\right)^{-1}\nonumber \\
	&~~\times \left(|\alpha_k|^2\left(\mu_k^{\star}x_k^{(j)}\right)^H \bm{q}_k + \dfrac{\mu_k^\star \gamma^r p^c }{1+\mu_k^\star \gamma^r p^c} \rho\lambda_{k,2}^* \bm{f}\right),\\
	x_k^{\star} & = \left(\bm{w}_k^\star\right)^H \bm{q}_k - \rho\lambda_{k,1}+\mu_k|\alpha_k|^2 x_k^{(j)},\\
	y_k^{\star} & = \dfrac{\left(\bm{w}_k^\star\right)^H \bm{f} - \rho\lambda_{k,2}}{1+\mu_k^\star \gamma^r p^c},
\end{align}
where $\mu_k^{\star}$ can be obtained when the equality is achieved  in the constraint (\ref{con:radar_sin3}) via the bisection search.

\subsection{Optimal Solution to Subproblem  (\ref{pb:u})}
The Lagrange function of subproblem (\ref{pb:u}) can be expressed as 
\begin{align}
\mathcal{L}=&|v|^2 \sum_{k=1}^K\left|\bm{p}^H\bm{u}_k\right|^2  + \dfrac{1}{2\rho}\sum_{k=1}^K |  \bm{h}_k^H \bm{u}_k-x_k -\rho\lambda_{k,1}|^2  \nonumber\\
&+\tilde{\lambda}\left(\sum_{k=1}^K||\bm{u}_k||^2 -P_{\max}\right),
\end{align}
where $\tilde{\lambda} \ge 0$ is the Lagrange multiplier for constraint (\ref{con:o_radar_power}). 
By examining the KKT conditions, we obtain
\begin{align} 
\dfrac{\partial \mathcal{L}}{\partial \bm{u}_k} =& 2|v|^2\left(\bm{p}^H\bm{u}_k\right) \bm{p}    + 2\tilde{\lambda}\bm{u}_k \nonumber \\
&+ \dfrac{1}{\rho} \left(  \bm{h}_k^H \bm{u}_k-x_k -\rho\lambda_{k,1}\right) \bm{h}_k= 0,~\forall k,\label{kkt:u_1}
\end{align}
\begin{equation}\label{kkt:u_2}
\tilde{\lambda}\left(\sum_{k=1}^K||\bm{u}_k||^2 -P_{\max}\right) = 0.
\end{equation}
Then, we can discuss the solution based on the value of $\tilde{\lambda}^{\star}$ as follows.

\textbf{Case 1}: If $\tilde{\lambda}^{\star}=0$, the optimal solution  satisfies that $\bm{p}^H\bm{u}_k^{\star}=0$, $\bm{h}_k^H \bm{u}_k^{\star}-x_k -\rho\lambda_{k,1}=0$, and $\sum_{k=1}^K||\bm{u}^{\star}_k||^2 \le P_{\max}$. Therefore, the optimal solution can be given by 
\begin{equation}
\bm{u}_k^{\star}\!=\!\dfrac{x_k+\rho \lambda_{k,1}}{||\bm{p}||^2||\bm{h}_k||^2 \!-\! |\bm{p}^H\bm{h}_k|^2}\left(||\bm{p}||^2\bm{h}_k-\bm{p}^H\bm{h}_k\bm{p}\right), \forall k.
\end{equation}

\textbf{Case 2}: If $\tilde{\lambda}^{\star}\neq 0$, according to (\ref{kkt:u_1}) and (\ref{kkt:u_2}), the optimal solution can be given by
\begin{equation}
\!\bm{u}_k^{\star}\!=\!\left(v^2\bm{p}\bm{p}^H\!+\dfrac{1}{2\rho}\bm{h}_k\bm{h}_k^H\!+\lambda^{\star}\bm{I}\right)^{-1}\!\!\dfrac{1}{2\rho}(x_k+\rho\lambda_{k,1})\bm{h}_k,
\end{equation}
where $\tilde{\lambda}^{\star}$  satisfies $\sum_{k=1}^K||\bm{u}_k||^2 =P_{\max}$ and can be obtained  via the bisection search.

\section{Proof of Lemma \ref{thm:opt}} \label{proof:opt}
First of all, we can optimize $\bm{w}_k$ since  it only appears in the constraint (\ref{con:o_radar_sinr}) and the optimal solution $\bm{w}_k^{R,\star}$ can be derived by utilizing the Rayleigh quotient maximization \cite{Rayleigh}, as 
\begin{equation}
	\bm{w}_k^{R,\star} = \left(\sigma^2 \bm{I} + p^c \hat{\bm{h}}_{ts}\hat{\bm{h}}_{ts}^H\right)^{-1}\bm{a}(\theta_k)\bm{a}^T(\theta_k)\bm{u}_k.
\end{equation}
With $\bm{w}_k^{R,\star}$, the problem (\ref{pb:o}) can be  equivalently transformed into the following problem:
\begin{subequations}\label{pb:non-2}
	\begin{eqnarray}
		& \max\limits_{\bm{u}_k} &   \sum_k\left|\hat{\bm{h}}^H_{sr}\bm{u}_k\right|^2,\\
		&\text{s.t.}&  {\left| \bm{a}^T(\theta_k)\bm{u}_k\right|^2} \ge \hat{\gamma}^r_k,~\forall k,\\
		&	&\sum_k\left|\left|\bm{u}_k\right|\right|^2 \le P_{\max}.
	\end{eqnarray}
\end{subequations}
Then, we can rewrite $\bm{u}_k$ as the linear combination of the $\bm{a}(\theta_k)$, $\bm{e}_{k}$, and $\bm{r}_{k}$, that is 
\begin{equation}
	\bm{u}_k = \eta_{k,1}\bm{a}^*(\theta_k) + \eta_{k,2}\bm{e}_{k} + \bm{r}_{k},
\end{equation}
where  $\bm{r}_{k}^H$ satisfies $\bm{r}_{k}^H \bm{a}(\theta_k)^* =0$, $\bm{r}_{k}^H \bm{e}_{k} =0$. Problem (\ref{pb:non-2}) can be simplified to
\begin{subequations}\label{pb:non-3}
	\begin{eqnarray}
		& \max\limits_{\{\eta_{k,1}, \eta_{k,2}\}} &   \sum_k\left|\hat{\bm{h}}^H_{sr}\bm{a}^*(\theta_k) \eta_{k,1} + \hat{\bm{h}}^H_{sr}\bm{e}_{k} \eta_{k,2} \right|^2,\\
		&\text{s.t.}&  {\left| \eta_{k,1}\right|^2} \ge \hat{\gamma}^r_k,~\forall k,\\
		&	&\sum_k\left(\left| \eta_{k,1}\right|^2+\left| \eta_{k,2}\right|^2 \right) \le P_{\max}. \label{con:non-3-energy}
	\end{eqnarray}
\end{subequations}
By  applying the method of Lagrange multiplier and  KKT conditions, we can obtain the following optimal solution to problem (\ref{pb:non-3})
\begin{align}
	\eta_{k,1}^\star  &= \sqrt{\hat{\gamma}_k},  \\
	\eta_{k,2}^\star  &= \left\{
	\begin{array}{ll}
	    -\dfrac{\hat{\bm{h}}_{sr}^T\bm{e}^*_{k} \hat{\bm{h}}_{sr}^H\bm{a}^*(\theta_k)}{ |\hat{\bm{h}}_{sr}^T\bm{e}^*_{k}|^2+ \hat{\lambda}^{\star}}	\sqrt{\hat{\gamma}_k}, & \text{if } |\hat{\bm{h}}_{sr}^T\bm{e}^*_{k}| \neq 0,\\
	     0, &  \text{if } |\hat{\bm{h}}_{sr}^T\bm{e}^*_{k}| = 0,
	\end{array}\right.~~\forall k,
\end{align}
where $\hat{\lambda}^{\star}\ge 0$ is the optimal Lagrange multiplier for constraint (\ref{con:non-3-energy}) and it satisfies  the  complementary slackness condition
\begin{equation}
	\hat{\lambda}^\star \left(\sum_{k=1}^K \left(\left| \eta_{k,1}^\star\right|^2+\left| \eta_{k,2}^\star\right|^2 \right) - P_{\max}\right) = 0.
\end{equation}
Based on the above, the optimal solution can be summarized in Lemma \ref{thm:opt}.

\section{Proof of Theorem \ref{thm:power}} \label{proof:power}
The interference from the radar to the communication receiver is given by 
\begin{align}
	&\sum_{k=1}^K\left|\hat{\bm{h}}^H_{sr}\bm{a}^*(\theta_k) \eta_{k,1} + \hat{\bm{h}}^H_{sr}\bm{e}_{k} \eta_{k,2} \right|^2 \nonumber\\
	=&\sum_k \hat{\gamma}_k \left|\hat{\bm{h}}^H_{sr}\bm{a}^*(\theta_k) \right|^2 \left|1-\dfrac{|\hat{\bm{h}}_{sr}^T\bm{e}^*_{k}|^2}{ |\hat{\bm{h}}_{sr}^T\bm{e}^*_{k}|^2 + \hat{\lambda}^{\star}} \right|^2,
\end{align}
which increases with $\hat{\lambda}^{\star}$. Meanwhile, the   complementary slackness condition for  $\hat{\lambda}^{\star}$ can be rewritten as 
\begin{equation}
	\hat{\lambda}^\star \left(\sum_{k=1}^K\hat{\gamma}_k\left(  1 +\dfrac{\left|\hat{\bm{h}}_{sr}^T\bm{e}^*_{k} \hat{\bm{h}}_{sr}^H\bm{a}^*(\theta_k)\right|^2}{ \left(|\hat{\bm{h}}_{sr}^T\bm{e}^*_{k}|^2+ \hat{\lambda}^{\star}\right)^2}	 \right) - P_{\max}\right) = 0.
\end{equation}
From the above equation, we can find that as $P_{\max}$ increases, $\hat{\lambda}^*$ decreases to satisfy to the complementary slackness condition, and $\hat{\lambda}^*$ becomes zero when $P_{\max}$ is large enough.  As a result, the interference from the radar to the communication receiver decreases to zero with $P_{\max}$. Then, we can conclude the relationship between the communication SINR and the radar power budget as shown in Theorem \ref{thm:power}.

\end{document}